\documentclass[final,epsfig,amsfont]{article}
\usepackage{latexsym}
\usepackage[dvips]{graphicx}

\newcommand{\eeq}{\end{equation}}
\newcommand{\beq}{\begin{equation}}
\newcommand{\nuq}[1]{\label{#1} \eeq}
\newcommand{\eqa}[1]{eq. (\ref{#1})}

\begin{document}
\title{Quantum Entropies and Decoherence for the Multiparticle Quantum Arnol'd Cat}

\date{}
\author{ Giorgio Mantica  \\ {\em Center for
Non-linear and Complex Systems}\\ Universit\`a dell'Insubria, Via
Valleggio 11, Como, ITALY \\ and INDAM, GNFM, \\ and  I.N.F.N. sezione di Milano.
}
\maketitle

\begin{abstract}
I study the scaling behavior in the physical parameters of dynamical entropies, classical and quantum, in a model of collision induced decoherence in a chaotic system. The treatment is fully canonical and no approximations are involved or infinite limits taken. I present this model in order to clarify my views on the debate about the nature, definition, and relevance of quantum chaos.
\end{abstract}

\vspace{.1in}
\begin{center}
{\it Dedicated to Giulio Casati on his eightieth }
\end{center}
\vspace{.1in}
\section{Preface: Forty years with Giulio Casati}

Forty years ago, in December 1982, I defended my master thesis in Pavia under the guidance of Italo Guarneri. He had introduced me to his coworker from the university of Milano, Giulio Casati, who in the same days was turning forty. Ever since, my scientific career has been marked by this encounter, in repeated periods of proximity and elongation. After a brief permanence in Milano, Giulio managed to put me in contact with Joe Ford, with whom I was a postdoc in Atlanta for many years, before taking a three-years turn to France with Daniel Bessis, a fourth important figure in my education.

\begin{figure}
   \begin{center}
     \includegraphics[height=40mm, angle=0]{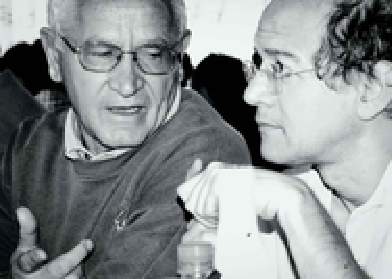}
  \end{center}
  \caption{Como, early 90's. Picture by Felix Izrailev}
  \label{fig-foto}
\end{figure}

But now back to Giulio: while I was in Paris, using his unrivalled political ability, he was setting up nothing less than a new university, in Como, where I later joined him (Figure \ref{fig-foto}) and Italo Guarneri in what at the beginning was a heroic endeavor: we shared a single office with a modem by which I connected to computers in Paris and Milano. We worked hard to establish the course in physics---later also in mathematics---and we brought an intense activity to the newly instituted Center for Non-linear and Complex Systems. We organized conferences, workshops; we invited many visitors, hosted postdocs and graduate students. Focus of this activity was chaos in classical and quantum systems, and also orthogonal polynomials of singular measures and their relevance for the phenomenon of quantum intermittency \cite{fbfunctions}, on which I will not dwell here.

\begin{figure}
   \begin{center}
     \includegraphics[height=40mm, angle=0]{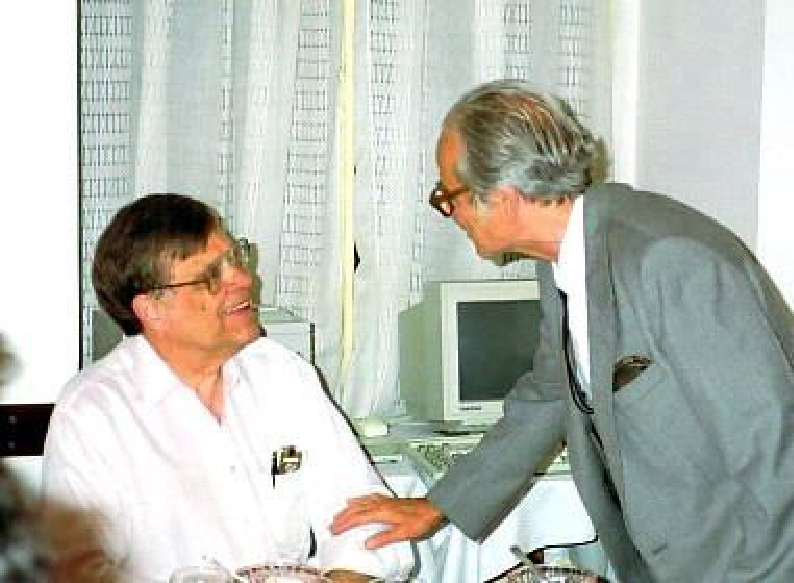}
  \end{center}
  \caption{Joe Ford and Boris Chirikov. Como, early 90's. Picture by Felix Izrailev}
  \label{fig-joe}
\end{figure}

In this paper I will instead present a reflection on the problem of quantum chaos and the quantum--classical correspondence. Notwithstanding the dedicatory goal, I will not hide the differences between my views and those of Giulio \cite{giuli,giulii,giubor,casachir1,casachir2}, the {\em extremist of the middle}, as Joe Ford jokingly called him, not without a hint of affection. The nickname implied that such views, Giulio's, were widely shared. Indeed, to the question {\em Does chaos pose a problem to quantum mechanics?} few would answer in the positive, and many would quote the notions of quantum chaology \cite{berry1,berry2} (Berry) or quantum pseudo-chaos \cite{boris1,viva} (Chirikov). Joe Ford  was convinced of the contrary and with some differences I shared his views \cite{physd} (but compare the different perspective of \cite{natoasi} and \cite{joeamj}). 
This celebration is now the occasion for me to put on the record what I have learnt in these years and what I believe is still to be investigated.

\section{Introduction and Summary of the Paper.}

As I mentioned, Giulio introduced me to Joe Ford, who had learnt the basic tenets of algorithmic complexity theory \cite{joe} from his colleague and friend Boris Chirikov (Figure \ref{fig-joe}) from the Soviet Union, where this theory had flourished thanks to the school of Kolmogorov. It leaves a sour taste to remember the many liaisons that ensued, in a moment when we fear for many dear friends and colleagues. This theory quantifies the amount of information contained in a sequence of symbols \cite{kolcomp1,kolcomp2,kolcomp3,kolcomp4}. Now, symbolic dynamics is a tool to translate motions into sequence of symbols, in such a way that one can equate chaotic trajectories of a system to random symbolic sequences---roughly speaking, sequences that cannot be significantly compressed without losing information.
In this perspective, chaos is the uninterrupted production of information by the motion of a dynamical system, just in the same way as a stochastic process could do; hence, in Ford's wording \cite{joewhat}, {\em chaos is deterministic randomness}. In a historical/theoretical introduction to classical chaos coauthored with Giulio for the Journal of the Italian Physical Society \cite{miogdf} we adopted this perspective:  that instability of trajectories, transverse homoclinic intersections, horseshoes are the dynamical {\em mechanisms} by which chaos {\em arises}, but not its {\em essence}, which is the algorithmic complexity of symbolic trajectories.

Similarly, in quantum dynamics, a variety of indicators have been introduced to unveil the mechanisms behind instability--or its lack. It is clearly not my aim here to list all contributions, early or recent, to this research, but just to name a few, for the sake of argument. Phase space representations, like the Wigner function, have been employed since long for comparison with classical dynamics, as in \cite{andrey1,andrey2}. In \cite{physd} this analysis was instrumental to find the eigenfunctions of the evolution operator and to expose the exponential propagation of mode number in the classical Liouville evolution, which impairs stability and practical reversibility, a technique later revisited by Benenti and Casati \cite{giuli}. Quantum Lyapunov exponents have also been defined, see {\em e.g.} in \cite{majew,vilela} and more recently much attention has been devoted to the out of time autocorrelation functions, see the paper by Kurchan in this volume \cite{jorge} and references therein.

At a more general level than these mechanisms there is the study of the algorithmic complexity of the quantum motion. Many years ago, Berry and Hannay quantized the Arnol'd cat map \cite{avez} via semiclassical means \cite{berry}. Ford, myself and Ristow \cite{physd} developed a fully Hamiltonian quantization and set out to estimate the algorithmic complexity of the evolution, finding that it grows like the logarithm of the dimension of the Hilbert space of the system plus the logarithm of the time--length of the evolution. Prior to saturation of such upper bounds (see the description in Sect. \ref{sec-af}) the quantum behavior reproduces the classical, as it has been repeatedly reported using the other indicators mentioned above. Yet, it was recognised that the {\em scaling} of the range of such {\em correspondence} with the physical parameters is fully unsatisfactory to justify a sensible approach to the classical limit (a sort of {\em logarithmic catastrophe} \cite{natoasi}).

As it is well known, a possible solution to this problem has been proposed, at the end of the last century, \cite{ian1,halli,ian2,zurpaz1,zurpaz2,zurek0,monpaz,marcos} by considering that systems are coupled to an external environment. I believe that this approach still need to be completed, or at least polished, in two respects.
Firstly, by casting it in a fully Hamiltonian approach, as I shall do by introducing a model system of collision-induced decoherence that avoids recurse to mean field approximations or infinite baths of oscillators, by which non-dynamical randomness might creep in uncontrolled.
Secondly, by framing it into an information theoretical perspective, employing the so--called Alicky-Fannes (AF) entropy \cite{alifa,alik}, which offers computational advantages and avoids some of the difficulties of algorithmic complexity. It must be noted that quantum dynamical entropies, generalizing the classical KS notion \cite{katok} were already developed in the early days of ``quantum chaos'' \cite{kos1,pechu,michael}.

The scheme of this paper is as follows: in the next section I briefly recall the notion of dynamical entropy.
Section \ref{sec-af} discusses the importance of dynamical entropies in the notion of chaos and the fundamental problem that one encounters in taking the classical limit of chaotic quantum systems. Section \ref{sec-cat} is a review of the quantum version of the classical Arnol'd cat map \cite{avez}, as the motion of a {\em single} kicked particle, so to open the way to a {\em multi} particle generalization. The Alicki-Fannes entropy of such single particle system has been computed in a series of papers. The formalism is presented in Section \ref{sec-afcat}, again in view of its generalization to the multiparticle Arnol'd cat \cite{gattosib}, described in Section \ref{sec-multicat}. This generalization aims at introducing in a fully controllable, dynamical way the concept of decoherence induced by scattering by light particles, following the paradigm of Joos and Zeh \cite{joos}. In a small aside, Section \ref{sec-von}, I briefly describe recent results on the {\em Von Neumann} entropy of the multi-particle Arnol'd cat \cite{condmat}, which confirm its suitability as model of quantum decoherence. In Section \ref{sec-multiaf} I return to the AF entropy, which in my view is theoretically more relevant than Von Neumann's, to the same degree as it is harder to treat theoretically and compute numerically. I explain how a combined limit in the number of particles and their mass {\em could} provide a physically sound means to understand the classical limit. Final comments are presented in the conclusions.

\section{Dynamical Entropy, Classical and Quantum}
\label{sec-entropies}

Consider a classical dynamical system \cite{avez,katok} consisting of a phase-space $X$ on which a transformation $\phi$ acts in such a way to preserve a measure $\mu$ defined on a suitable sigma-algebra of subsets of $X$: $\mu(\phi^{-1}(A)) = \mu(A)$ for any $A$ in the sigma-algebra. The map $\phi$ generates the dynamics, that we assume to evolve in discrete time instants, denoted by the symbol $n$ throughout the paper. Consider a finite partition $\cal P$ of $X$ by measurable subsets $P_{l}$, $l=0,\ldots,L-1$ , $\bigcup_l P_l = X$. The Shannon entropy of this partition, $S({\cal P})$ is defined as
\begin{equation}\label{ent1}
S({\cal P})=-\sum_{l=0}^{L-1} \mu(P_{l})\log\mu(P_{l}).
\end{equation}
This value is the average amount of information obtained by determining the element of the partition that contains a given point in phase space: this action can be physically interpreted as the result of a measurement experiment.

Next, consider the partition obtained by intersecting all elements of ${\cal P}$ and all their pre-images by the map $\phi$: indicate this new partition by
\beq
{\cal P} \vee \phi^{-1}({\cal P}) = \{ P \cap \phi^{-1}(Q), \mbox{ for } P, Q \in {\cal P} \},
\nuq{eq-part2}
where obviously some of the intersections at right hand side may be empty. If we label sets by the {\em alphabet} composed of the {\em symbols} $0$ to $L-1$, as in \eqa{ent1}, the elements of the partition in \eqa{eq-part2} are labeled as $P_{lk} = P_l \cap \phi^{-1}(P_k)$. These sets comprise all points that at time zero are located in $P_{l}$ and at the successive time instant in $P_{k}$. Therefore, the Shannon entropy of ${\cal P} \vee \phi^{-1}({\cal P})$ measures the average amount of information provided by knowledge of the {\em symbolic dynamics} of a point over two time instants.

In the lore, a {\em word} is a sequence of symbols $\mathbf{\sigma} = \sigma_0, \sigma_1, \ldots, \sigma_{n-1}$ in the alphabet. We use the notation $|\sigma|$ for the length of a word. Let $\Sigma_n$ be the set of all words of length $n$ in the alphabet. Clearly, the cardinality of this set is $L^n$. Every word defines a {\em cylinder} $P_\mathbf{\sigma}$ in $X$: this is the set of all points $x$ for which  $\phi^i(x) \in P_{\sigma_{i}}$, for $i=0,\ldots,n-1$. The set of cylinders associated with words of length $n$ compose the partition $\vee_{i=0}^{n-1} \phi^{-i}({\cal P})$, whose entropy is therefore
\beq
  S({\cal P},n) = S(\vee_{i=0}^{n-1} \phi^{-i}({\cal P})) = - \sum_{\mathbf{\sigma} \in \Sigma_n} \mu(P_\mathbf{\sigma}) \log(\mu(P_\mathbf{\sigma})).
\nuq{eq-entr2}
Consider the difference $S({\cal P},n+1)-S({\cal P},n)$. This quantifies the average amount of new information obtained by observing the state of the motion at time $n$, on top of what already obtained by the observations at the previous $n$ instants, from time $0$ to time $n-1$. It could also be described as the {\em surprise} that the dynamical evolution of the system is capable of producing at such time. It is then understandable that one can define as dynamical entropy the asymptotic rate of information production as
\beq
   h({\cal P}) = \lim_{n \to \infty} S({\cal P},n+1)-S({\cal P},n)
   = \lim_{n \to \infty} \frac{1}{n} {S({\cal P},n)}.
\nuq{eq-entr3}
This leads to the definition of the Kolmogorov Sinai entropy of the dynamical system as a supremum over partitions:
\beq
   h_{KS} (X,\phi,\mu,{\cal A}) =
   \sup \{ h({\cal P}), {\cal P} \mbox{ finite measurable partition of } X \}.
\nuq{eq-entr4}
Chaotic classical dynamical systems are then properly defined as those having positive KS entropy, and Brudno's theorem \cite{aleks} links this concept to positive algorithmic complexity.

Let us now turn to quantum dynamics for comparison. Robert Alicki and Mark Fannes have constructed a definition of entropy following the same steps as in the classical construction \cite{alifa,alik,benabook}. A quantum dynamical system is composed of a Hilbert space ${\cal H}$ and of a unitary evolution $U$ that leaves invariant a density matrix $\rho$.  The analogue of the partition ${\cal P}$ is a finite family of orthogonal projection operators $P_l$, from $l=0$ to $L-1$, such that $\sum_l P_l = I$, $I$ being the identity. We use the same symbol $P_l$ for both the classical and the quantum case, since this highlights the analogy.

For illustration, suppose that $\rho = | \psi \! >< \! \psi |$, even if generically this is not invariant. The wave function $| \psi \! >$ represents the initial state of the system. Projecting $|\psi \! >$ by $P_l$ is the analogue of the classical {\em measurement} of the partition element which the dynamical state belongs to. In quantum dynamics, the square modulus $\|P_l|\psi>\|^2$  is the probability that the result of this {measurement} yields the value $l$. Clearly, because of normalization, $\sum_l \|P_l|\psi \! >\|^2 = 1$. Following suit, the analogues of the elements of ${\cal P} \vee \phi^{-1}({\cal P})$ are
$
  P_{\sigma_{1}} U P_{\sigma_0} |\psi\! >
$,
which shows that the classical action of $\phi^{-1}$ is replaced by the quantum unitary evolution operator $U$. This leads to define the quantum dynamical partitions
\beq
\vee_{i=0}^{n-1} U^{i}({\cal P}) = \{ P_{\sigma_{n-1}} U \circ P_{\sigma_{n-2}}  U \circ \cdots \circ P_{\sigma_{1}}  U \circ P_{\sigma_0}, \; \mathbf{\sigma} \in \Sigma_n \} =
  \{ P_\sigma, \; \mathbf{\sigma} \in \Sigma_n \}.
\nuq{eq-veq1}
We have used the notation $P_{\mathbf{\sigma}}$ for the operators in the above set. Then, the quantum analogue of $\mu(P_\sigma)$ takes the form of the trace of a product of operators:
\beq
   \mu(P_\sigma) = \mbox{Tr} ( P^\dagger_\sigma \rho P_\sigma ).
\nuq{eq-trace1}
Alas, it is not possible to use this quantity to define a dynamical entropy using \eqa{eq-entr2}. In fact, suppose again for simplicity that $\rho = | \psi \! >< \! \psi |$. Then
\beq
   \mu(P_\sigma) = \| P_\sigma | \psi \! >\|^2.
\nuq{eq-trace2}
While it is true that $\sum_\sigma \mu(P_\sigma) = 1$, observe that \eqa{eq-trace2} can be rewritten as follows:
\beq
\| P_\sigma | \psi \! >\|^2 = <\! \psi P_\sigma | P_\sigma \psi \! >.
\nuq{eq-trace2a}
The word ${\mathbf \sigma}$ appears twice in the above equation. Replace the first appearance by a different word ${\mathbf \theta} \in \Sigma_n$, to get
\begin{eqnarray*}
 <\! \psi P_\theta | P_\sigma \psi \! > =
 <\! \psi  | P^\dagger_\theta P_\sigma \psi \! > =
<\! \psi  | P^\dagger_\theta P_{\sigma_{n-1}} U \circ
\cdots \circ P_{\sigma_{1}}  U \circ P_{\sigma_0} \psi \! > = \\
= <\! \psi  | P_{\theta_{0}} \circ U^{-1} P_{\theta_{1}} \circ \cdots \circ U^{-1} P_{\theta_{n-1}} \circ P_{\sigma_{n-1}} U \circ
\cdots \circ P_{\sigma_{1}}  U \circ P_{\sigma_0} \psi \! >.
\label{eq-trace2b}
\end{eqnarray*}
It helps intuition to read the last line from right to left: in sequence, the initial function $|\psi \! >$ is projected by $P_{\sigma_0}$, acted upon by $U$, projected by $P_{\sigma_1}$, acted upon by $U$, and so on until it is projected by $P_{\sigma_{n-1}}$. Then, the adjoint of  $P_{\mathbf \theta}$ operates by projecting by $P_{\theta_{n-1}}$, acting by the {\em inverse} of $U$ (that is, by reversing time), projecting by $P_{\theta_{n-2}}$, acting by  $U^{-1}$ and so on until the projection by $P_{\theta_0}$ and finally taking the scalar product with the initial function $|\psi \! >$.

Clearly, the result of this operation is null unless $\sigma_{n-1}=\theta_{n-1}$. In the classical case where $\phi^{-1}$ takes the place of $U$ the result is also null unless the {\em full} word $\mathbf \theta$ coincides with $\sigma$. In fact, it is enlightening to consider the above process as a sequence of many--slits experiments: a dynamical trajectory must go through the slits $\sigma_0$ to $\sigma_{n-1}$. Then, time is reversed and it is required to pass from the slits $\theta_{n-1}$ to $\theta_0$. Since trajectories in classical dynamical systems are time reversible, this is only possible if $\mathbf{\theta} = \mathbf{\sigma}$.
On the contrary, in quantum dynamics two different {\em paths} from $\sigma_0=\theta_0$ to $\sigma_{n-1}=\theta_{n-1}$, the end points being fixed, may {\em not} be orthogonal because of {\em quantum interference}:
 $<\! \psi P_\theta | P_\sigma \psi \! > = <\! \psi_\theta | \psi_\sigma \! > \neq 0$,
where we have introduced the notation $|\psi_\sigma \! > = P_\sigma \psi \! >$.

If we now take for $\rho$ an invariant density matrix (which in this paper will be the normalized identity matrix $I$), we define the
{\em decoherence matrix} $D$ with entries $D_{\theta,\sigma}(n)$
\beq
  D_{\theta,\sigma}(n) =
   \mbox{Tr} ( P^\dagger_\theta  \rho  P_\sigma ) =\mbox{Tr} ( P^\dagger_\theta   P_\sigma ) =
  \frac{1}{\cal N} \sum_{j=1}^{\cal N} <\! \psi^j_\theta | \psi^j_\sigma \! >,
\nuq{eq-trace1b}
where ${\cal N}$ is the dimension of the Hilbert space and $\psi^j \! >$ is an orthonormal basis of ${\cal H}$. In this paper I shall consider a notable example of such finite-dimensional Hilbert space.
The decoherence matrix is a fundamental object of the {\em coherent histories} formalism \cite{histo1,histo2,histo3,histo4,histo5,histo6}.
Observe that $D(n)$ is an $L^n \times L^n$ square matrix, Hermitean, of unit-trace and non-negative. In the classical case, this matrix is diagonal. In the quantum case, one defines the partition entropy as
\begin{equation}
   S({\cal H}, U, {\cal P}, n) = - Tr ( D(n) \log D(n))
  \label{proj3}
 \end{equation}
and can therefore proceed to define the AF entropy operating as in equations (\ref{eq-entr3}), (\ref{eq-entr4}) \cite{benabook}. To suit our purposes we do not take the infinite $n$ limit, or the supremum over partitions, but rather consider the scaling behavior of the quantity $S({\cal H}, U, {\cal P}, n)$, concisely denoted $S(n)$ that we will call AF entropy for short (Shannon AF entropy at finite resolution and time being a more precise term) for physically accessible partitions, ranges of time, and physical parameters.

\section{Dynamical entropy and quantum chaos}
\label{sec-af}

On the same theoretical basis as in classical dynamics, we shall consider as chaotic a quantum system capable of producing new information at an inferiorly bounded rate indefinitely in time when measured by a sufficiently refined partition $\cal P$. For this, we adopt the AF quantum entropy defined in the previous section. It helps to remark that Benatti has derived a quantum version of the classical Brudno theorem, linking AF entropy and algorithmic complexity \cite{benatti}, so that the parallelism with classical dynamics is completed.

At a superficial inspection it may seem that there should be no limitation for the entropy $S({\cal H}, U, {\cal P}, n)$ in \eqa{proj3} to grow linearly with $n$: as mentioned, the matrix $D$ has size $L^n \times L^n$. Yet, thanks to a result in \cite{alik} the spectrum of non-zero eigenvalues of $D$ coincides with that of a second matrix, $\Omega(n)$, of size ${\cal N}^2 \times {\cal N}^2$ independent of $n$, of unit trace and Hermitean. This matrix is the basis of a technique to numerically {\em compute} $Tr [ D(n) \log D(n)]$ combining the theory of orthogonal polynomials with a sampling method from lattice gauge theory \cite{etna}. It appears to be the only method proposed so far capable of attacking non--trivial Hilbert space and word lengths dimension.
While this technique is rather complicated, one consequence of its is immediate: $S({\cal H}, U, {\cal P}, n)$ is the Shannon entropy of a probability distribution over ${\cal N}^2$ discrete states, whose number is {\em independent} of $n$. Therefore, it is bounded by $2 \log {\cal N}$ and the limit in \eqa{eq-entr3} is forcefully null.

Why this is a serious problem can be easily understood: consider a classical dynamical system that can be {\em quantized} via the standard procedure of replacing classical variables by quantum operators. Further assume that the resulting Hilbert space has finite dimension ${\cal N}$. Consider a classical partition ${\cal P}$ that corresponds to a quantum family of projectors. One can compute the classical probabilities $\mu({\cal P}_{\mathbf \sigma})$ as well as the eigenvalues of the quantum decoherence matrix $D(n)$ and then compare the associated Shannon entropies at variable duration $n$ of the experiment. The {\em correspondence principle} leads us to expect that the two results be comparable, in the {\em classical} limit. Yet, recall that the quantum system is endowed with a finite {\em information reservoir}. In the initial instants of the evolution the classical system is outputting information and the quantum analogue follows suit, keeping pace with the classical.  Yet, this can last as long as the quantum system has not fully exhausted its information reservoir.

Suppose now that the classical system is {\em not} chaotic, in the sense that $S({\cal P},n)$ grows {\em less} than linearly, say logarithmically, with $n$. In this case, the value of $n$ at which the quantum dynamics can provide the same information output as the classical is roughly proportional to $\cal N$, and therefore sufficiently large to justify a physical relevance of correspondence.
On the contrary, when the classical dynamics is chaotic, {\em i.e.} $S({\cal P},n)\sim h({\cal P}) \; n$, the finite quantum information reservoir is eaten away at a rate sustainable only up to $n \sim \log {\cal N}$, after which the two evolutions qualitatively and quantitatively differ. This is in essence the logarithmic catastrophe in the correspondence principle mentioned before.

From a physical point of view, the last estimate corresponds to what has been defined the Zaslavsky time scale \cite{zas} (also termed Ehrenfest time scale \cite{dima}) $\tau = \log ({\cal I}/\hbar)$, where ${\cal I}$ is a typical action of the system, a fact that yields the analogy ${\cal N} \sim ({\cal I}/\hbar)$ and prompts matter for thought. 
In this paper, $\hbar$ will be constant, while the {\em mass} of a quantum particle will determine ${\cal N}$.

From a theoretical point of view, advocated by Berry among others \cite{berry1,berry2}, it is clear that dynamical entropies are properly defined only in the limit when $n$ tends to infinity, while the classical limit is ${\cal I}/\hbar \rightarrow \infty$ and the two limits do not commute. At the same time, it is apparent that this attitude ignores the unphysically low value of the Zaslavsky time for {\em e.g.} the chaotic three body problem. Moreover, as I have first suggested in \cite{natoasi}, rather than considering the limit for $n$ tending to infinity, one should study the {\em scaling relations} with respect to the physical parameters of the information--theoretical quantities (\ref{eq-entr3}) and (\ref{proj3}) in realistic ranges of the physical variables. In other words: will a sequence of physically realizable experiments at finite precision be able to reveal a significant difference among the two dynamics in the semiclassical region? A model of this situation is found in the dynamics of classical billiards \cite{algocom1}: a difference emerges clearly in the behavior of the finite time algorithmic complexity of symbolic trajectories, when polygonal tables approach chaotic or integrable curved billiards.

In the next section I briefly recap the details of a much investigated system, the quantum Arnol'd cat, where this situation emerges with clarity. This system is also particularly useful to study the arguably most significant way to explain the emergence of classical properties (among which chaos) in quantum dynamics: {\em decoherence}. In this approach, a quantum system is not isolated, but coupled to an {\em environment} that is responsible for the decay of the off-diagonal matrix elements of $D$. Yet, in the usual application of this approach the environment is mathematically represented by a random source, or an infinite bath of oscillators: in any case a system possessing an {\em infinite} amount of information, and not, to the best of my knowledge, by a quantum system with finite dimensional Hilbert space. In previous works \cite{gattosib} I have proposed a scheme by which the environment is composed of the tensor product of simple low--dimensional systems, so that the dimension of the full, purely quantum, isolated pair system--environment grows exponentially in the number of the elements of the latter and can therefore provide a physically reasonable solution to the logarithmic catastrophe described above. The multi-particle Arnol'd cat introduced in this paper is a simple model to implement this scheme.
Due to technical difficulties, both theoretical and numerical, much is still to be understood on the information flow in this system. 
But first, let us review the original quantum Arnol'd cat.

\section{Short review of the quantum Arnol'd cat}
\label{sec-cat}

Despite the fact that this quantum system has now become {\em classical} (pun), it is interesting to present a derivation that opens the way to its decoherent generalization.  We study a point particle of mass $M$ free to rotate on a ring, without gravity. This ring is a one--dimensional torus of length $L$ and position on it is $X \in [0,L)$. Most of the time, the particle evolves freely at constant velocity. At discrete times, it is subject to a periodic impulsive force, of period $T$, the relative Hamiltonian being:
\begin{equation}
 H_{cat}(X,Y,t) = \frac{Y^2}{2M} -  \alpha \frac{X^2}{2} \sum_{j=-\infty}^\infty \delta(t/T-j).
 \label{eq-ham1}
\end{equation}
$Y$ is the conjugate momentum to $X$ and $\alpha$ is a coupling constant. According to Hamilton's equations, the effect of the force is to instantaneously change the particle's momentum.

Let us show that $H_{cat}$ is the Hamiltonian that yields the Arnol'd cat mapping.
Consider the classical period evolution of the dynamical variables observed at instants immediately following the action of the impulsive force:
\begin{equation}
  \left\{
\begin{array}{l}
X \rightarrow  X +  \frac{T}{M} Y    \\
Y  \rightarrow Y (1+ \frac{\alpha T}{M}) + \alpha T X .
\end{array} \right.
\label{eq-ham2}
\end{equation}
Introduce the rescaled momentum $\tilde{Y}:=  \frac{T}{M} Y$, which has the dimensions of a length. The period evolution operator in the new variables is:
\begin{equation}
  \left\{
\begin{array}{l}
X \rightarrow  X +  \tilde{Y}  \\
\tilde{Y}  \rightarrow \frac{\alpha T^2}{M} X + (1+\frac{\alpha T^2}{M}) \tilde{Y} .
\end{array} \right.
\label{eq-ham3}
\end{equation}
Require that the rescaled momentum variable $\tilde{Y}$ also be periodic, of the same period $L$ as of $X$.
Choosing $L=1$ and $\frac{\alpha T^2}{M}$ to be an integer fulfills this requirement.
Finally, the Arnol'd cat mapping on the unit two--dimensional torus is obtained by further imposing that
\begin{equation}
\frac{\alpha T^2}{M}  = 1.
\label{eq-ham4}
\end{equation}

A rigorous mathematical framework for the quantum dynamics of this map can be found in \cite{isola}. I now follow the treatment in \cite{physd}, with results in agreement with the original work of Schwinger \cite{schw} on the quantum dynamics on a torus. Because of the $X$ periodicity the wave--functions takes the form
 \begin{equation}
  \psi (X) = \sum_k c_k \phi_k(X) = \sum_k c_k e^{i 2 \pi k X/L}.
  \label{eq-ham5}
\end{equation}
In the above, $\phi_k(X) := e^{i 2 \pi k X/L}$ are momentum eigenfunctions and $c_k$, with $k$ integer, are the expansion coefficients. Next, we consider periodicity in $Y$ with period $M L/T$.  The momentum operator is $\mathbf{Y} = -i \hbar \partial_X$, so that $\mathbf{Y} \phi_k(X) = \frac{k h}{L} \phi_k(X)$, $h$ is Planck's constant and therefore periodicity in momentum implies that
\begin{equation}
c_k = c_{k+{\cal N}}
\label{eq-ham6v}
\end{equation}
for any $k$, where ${\cal N}$ is an integer number:
\begin{equation}
\frac{M L^2}{T}  = {\cal N} h.
\label{eq-ham6}
\end{equation}
One parameter can be freely chosen and we set $L=1$, with no loss of generality.
One can explain the rationale behind \eqa{eq-ham6}: to embed quantum dynamics in a two dimensional torus of unit periodicity in the $X$ direction, the periodicity in $Y$ must be an integer multiple of the Planck constant.
Finally, using $c_k$ = $c_{k+{\cal N}}$ in \eqa{eq-ham5} implies that $\psi(X)$ takes the form of a periodic train of delta functions at the points $X_j$:
\beq
  X_j = \frac{j}{\cal {\cal N}}+s, \; j=0,\ldots,{\cal {\cal N}}-1
 \label{eq-qbigj}
 \end{equation}
where $s$ is an arbitrary constant shift. We shall use the notation $|j>$ for the delta at such positions.

The Hilbert space of the system is so isomorphic to $\mathbf{C}^{\cal {\cal N}}$. In the position representation wave functions can be written as
 \begin{equation}
  \psi (X_j) = \frac{1}{\sqrt{{\cal N}}} \sum_{k=0}^{{\cal N}-1} c_k \phi_k(X_j) =
  \frac{1}{\sqrt{{\cal N}}} \sum_{k=0}^{{\cal N}-1} c_k e^{-i 2 \pi k j/{\cal N}}
  \label{eq-ham5b}
\end{equation}
and the discrete Fourier transform maps the above to the momentum representation:

\begin{equation}
  c_k = \hat{\psi}(Y_k) = \frac{1}{\sqrt{{\cal N}}} \sum_{j=0}^{{\cal N}-1} \psi(X_j)e^{i 2 \pi k j /{\cal N}},
  \label{eq-ham5c}
\end{equation}
where $Y_k = k h$.

An important remark at this point is the following: via \eqa{eq-ham6} the dimension of the Hilbert space is directly proportional to the mass $M$ of the particle. Therefore, we can take the classical limit in its correct physical form, by keeping $\hbar$ to its real physical value and by considering a particle of larger and larger mass $M$.

Quantum motion induced by the Hamiltonian \eqa{eq-ham1} becomes a unitary operator in ${\cal H} = \mathbf{C}^{\cal N}$ that has been computed in \cite{physd}. In the position representation, where $\psi$ is the vector $\psi(X_j)$, $j=0,\ldots,{\cal N}$, the unitary evolution $U^{0}$ induced by the free rotation $\frac{Y^2}{2M}$ has matrix elements
\begin{equation}
U^{0}_{kl} = \frac{1}{\sqrt{{\cal N}}} e^{- (\pi i l^2 /{\cal N})} e^{2 \pi i kl /{\cal N}}.
  \label{eq-rota1}
 \end{equation}
The instantaneous  change in momentum induced by the kick is rendered by the operator $K$ with matrix elements:
\begin{equation}
K_{kl} = \frac{1}{\sqrt{{\cal N}}} e^{ i \pi l^2 / {\cal N}} \delta_{k,l},
  \label{eq-gatto1}
 \end{equation}
where $\delta_{k,l}$ is the Kronecker delta. To complete the review of the formulae in \cite{physd},
the quantum Arnol'd cat evolution operator is the product $U = K U^{0}$.

\section{The Alicki-Fannes entropy of the quantum Arnol'd cat}
\label{sec-afcat}

The theory of the Alicki-Fannes entropy can be applied to the dynamics of the quantum Arnol'd cat, using projections operators $P_l$ measuring the position coordinate only: letting $|j>$ for $j=0, \cdots, {\cal N}-1$ be the canonical basis vectors of $\mathbf{C}^{\cal {\cal N}}$ and letting for convenience
${\cal N} = 2^q$, $Q=2^{q-p}$ with $p$ and $q$ positive integers, $p<q$, we define
 \begin{equation}
  P_l = \sum_{j=l Q}^{(1+l)Q-1} |j><j|, \;\;\; l = 0,\ldots,2^{p}-1.
  \label{proj1}
 \end{equation}
For illustration, $P_0$ is the projection on the first $Q=2^{q-p}$ basis states, which corresponds to the classical projection on the phase-space set $0 \leq X < 1/2^p$, that is, the leftmost $1/2^p$ section of the unit two-dimensional torus (notice that all values of $Y$ are allowed).

\begin{figure}
\includegraphics[width=.6\textwidth, angle=270]{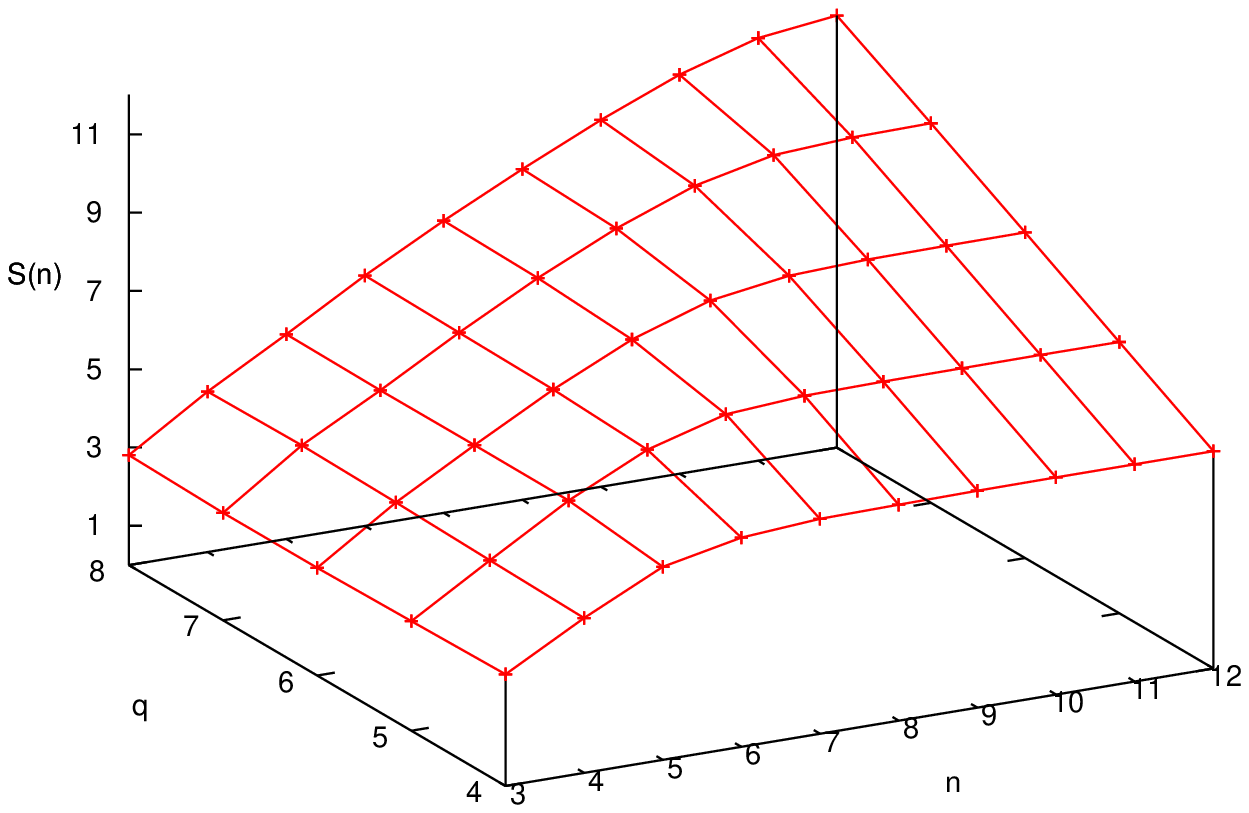}
\caption{AF entropy $S(n)$ as a function of $n$ and $q$.}
\label{fig-1}
\end{figure}

Even though the adaptation of the formalism of Section \ref{sec-entropies} to the case of the quantum Arnol'd cat is theoretically straightforward and physically transparent, its practical implementation, in terms of numerical analysis, is particularly challenging. Details of the sophisticated technique required for its implementation can be found in \cite{etna}. Following the approach presented in \cite{alik}, this technique permits to compute the entropy for arbitrarily long times avoiding the exponential increase of the dimension of the decoherence matrix. In Figure \ref{fig-1} the quantity $S(n)$ (shorthand notation for $S({\cal H}, U, {\cal P}, n)$) is plotted versus $n$ for different values of $q$ (recall, ${\cal N} = 2^q$) and $p=2$. We see that the bound $S(n) \leq 2 \log({\cal N})$ is approached at a rapid pace in $n$ for any finite value of $q$. Before this saturation, which happens when $n$ is proportional to $q$, that is $\log({\cal N})$, there is a small interval when $S(n)$ increases linearly with a slope dictated by the classical Lyapunov exponent. After this brief lapse of time, output of new information fades.

This is the logarithmic catastrophe described above. For indeed to {\em linearly} increase the length of this span of chaotic freedom one has to {\em exponentially} increase the dimension of phase space, {\em i.e.} the {\em mass} of the particle! It is clear that, while mathematically the classical complexity is fully regained in the limit, this is quite unsatisfactory from a physical viewpoint.

\begin{figure}
  \begin{center}
    \begin{tabular}{cc}
      \resizebox{80mm}{!}{\includegraphics[height=60mm, angle=-90]{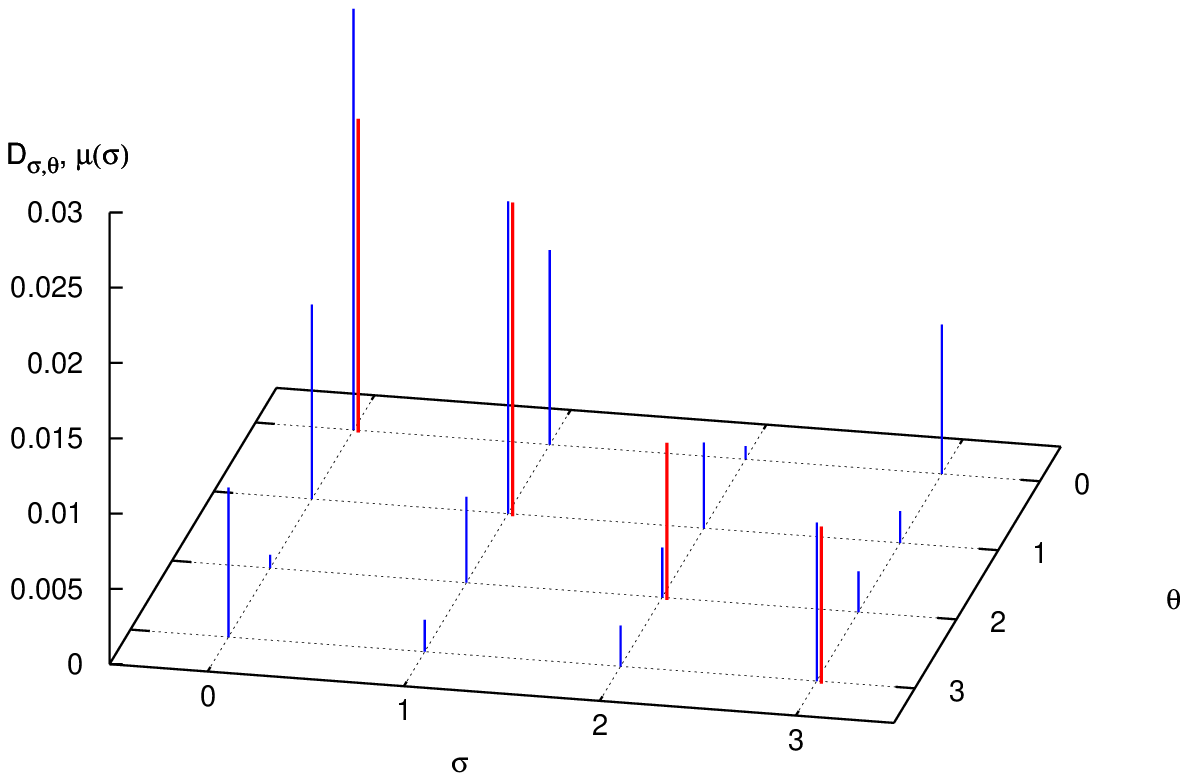}}   &
 \resizebox{80mm}{!}{\includegraphics[height=60mm, angle=-90]{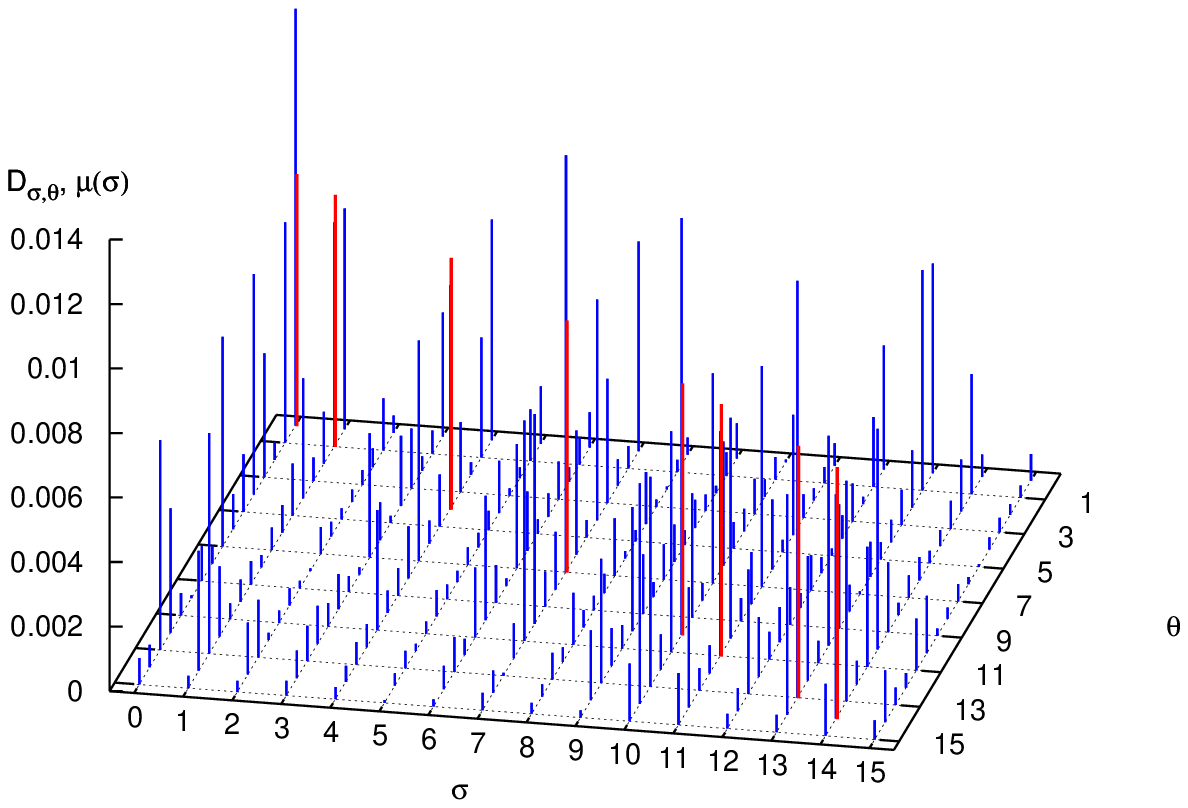}}  \\
    \end{tabular}

  \end{center}
  \caption{Absolute value of the decoherence matrix elements $D_{\sigma,\theta}$ (blue bars) and classical word probabilities $\mu(\sigma)$ (red bars, plotted on the diagonal for comparison), when $q=2$ so that ${\cal N} = 2^q = 4$, and $p=2$. In the left panel $|\sigma|=|\theta|=3$, in the right panel $|\sigma|=|\theta|=4$. See text for further detail.}
 \label{fig-deco00}\end{figure}

A detailed analysis of the decoherence matrix yielding these results further reveals their physical content.
In the numerical experiments that follow, we fix $p=2$, meaning that phase space and Hilbert space are partitioned in four sets (see \eqa{proj1} and discussion) and consequently the decoherence matrix is the direct sum of sixteen diagonal blocks, of which we consider the block $\sigma_0=\theta_0=0$, $\sigma_{n-1}=\theta_{n-1}=0$, the other blocks providing similar results. In Figure \ref{fig-deco00}, left panel, the matrix elements $|D_{\sigma,\theta}|$ are plotted as blue bars. The wordlength is three, so that interference between the paths $(0,\sigma_1,0)$ and $(0,\theta_1,0)$, as in a four-slit experiment, is clearly evident. Because of this, the diagonal values compare poorly with the classical probabilities $\mu(\sigma)$ (a shorthand notation for $\mu(P_\sigma)$), plotted as red bars on the diagonal. The forest of off-diagonal blue bars is even more evident when considering words of length four $(0,\sigma_1,\sigma_2,0)$ (there are 16 of them), in the right panel of Figure \ref{fig-deco00}.

\begin{figure}
  \begin{center}
    \begin{tabular}{cc}
      \resizebox{80mm}{!}{\includegraphics[height=60mm, angle=-90]{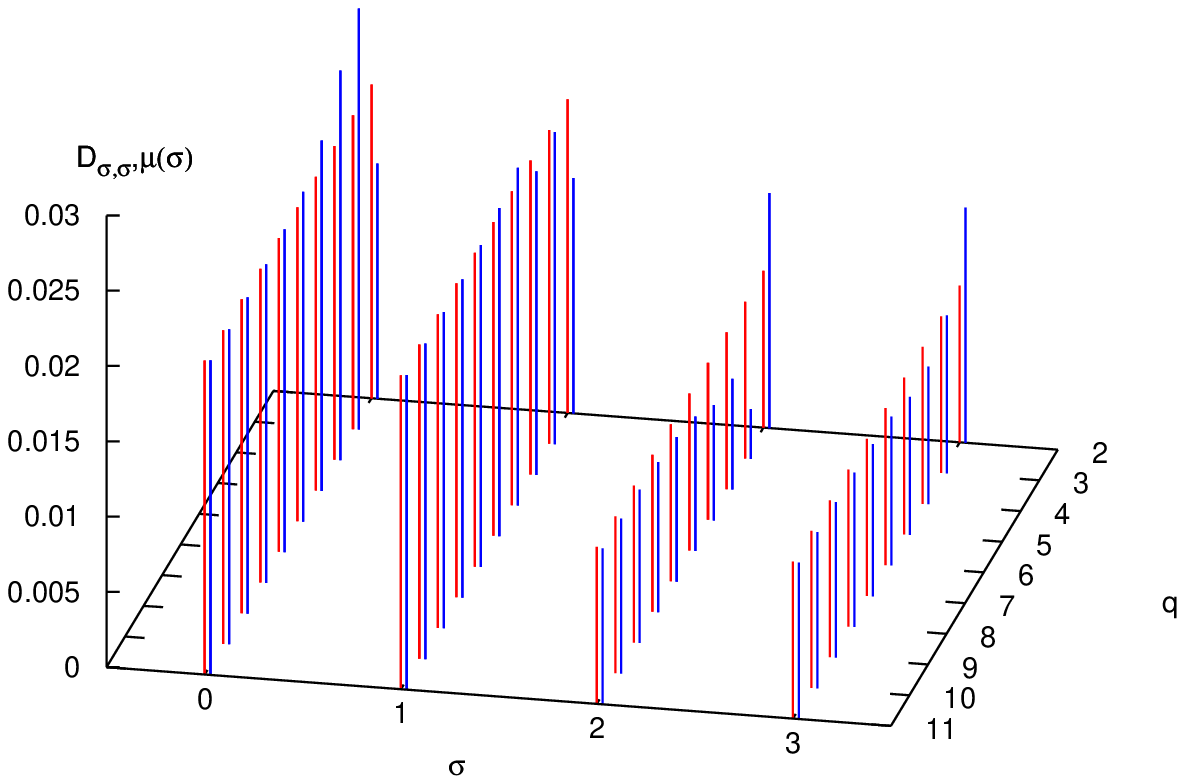}}   &
 \resizebox{80mm}{!}{\includegraphics[height=60mm, angle=-90]{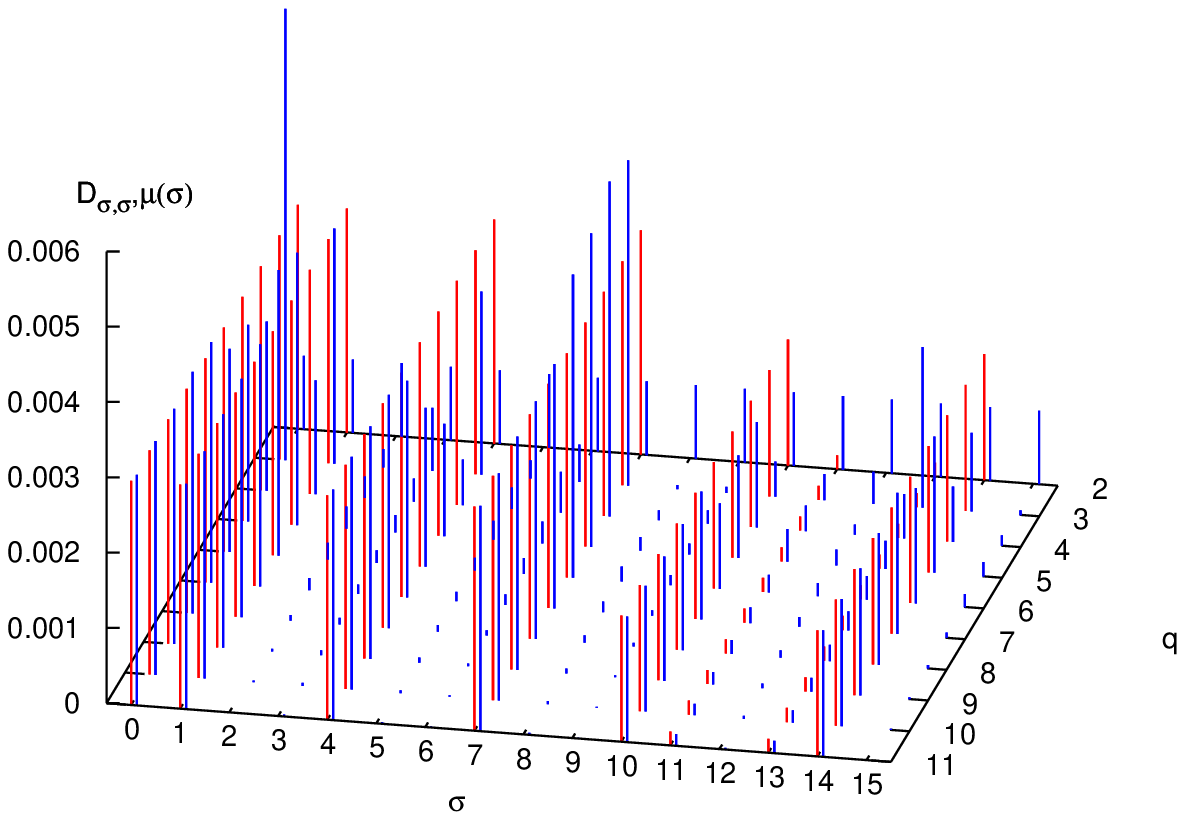}}  \\
    \end{tabular}

  \end{center}
  \caption{Absolute value of the diagonal decoherence matrix elements $D_{\sigma,\sigma}$ (blue bars) as function
   of $q$, compared with the classical word probabilities $\mu(\sigma)$ (red bars, obviously independent of $q$ and plotted sideways for comparison). In the left panel $|\sigma|=3$, in the right panel $|\sigma|=4$. See text for further detail.}
 \label{fig-deco01}\end{figure}

\begin{figure}
  \begin{center}
 \begin{tabular}{cc}
      \resizebox{70mm}{!}{\includegraphics[height=60mm, angle=-90]{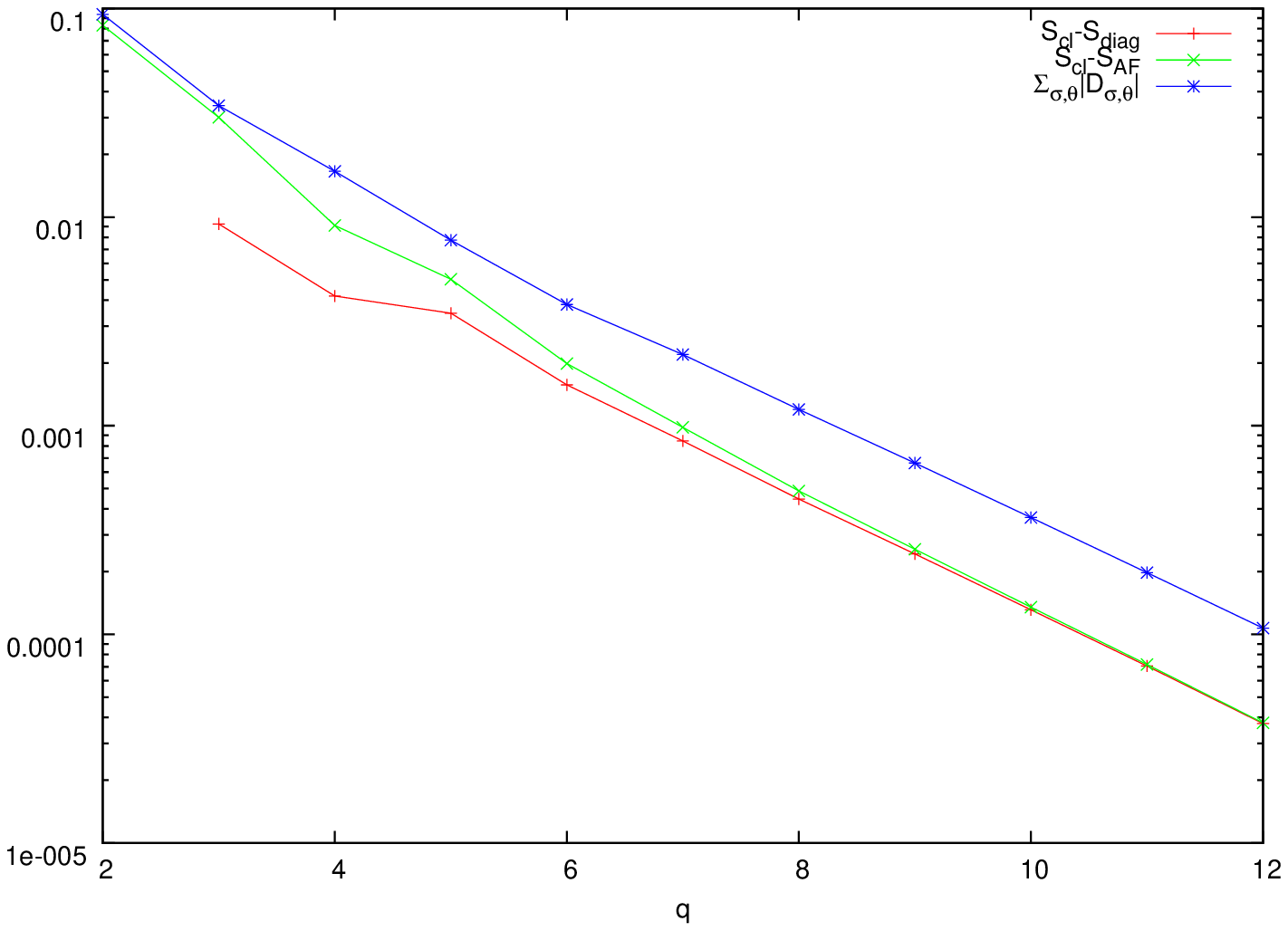}}   & \resizebox{80mm}{!}{\includegraphics[height=60mm, angle=-90]{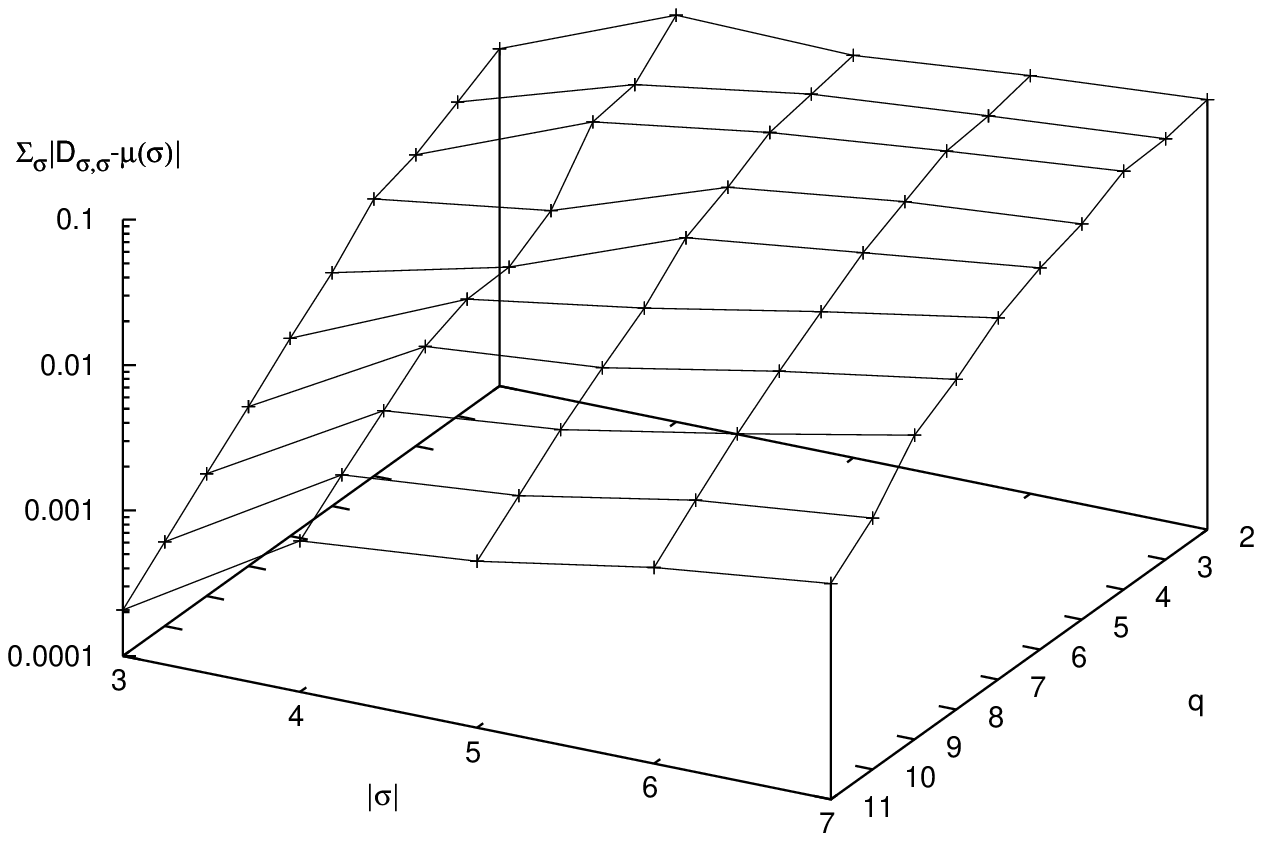}}
  \\
    \end{tabular}

  \end{center}
  \caption{Left panel: for $|\sigma|=3$, sum of the absolute values of the off-diagonal matrix elements $D_{\sigma,\theta}$ (blue curve), absolute difference between classical entropy $S_{cl}$ and {\em diagonal} entropy $S_{diag}$ (red curves), absolute difference between classical entropy $S_{cl}$ and AF entropy $S_{AF}$ (green curves). Right panel: symbolic distance $d_{\mbox{\tiny symb}} = \sum_\sigma |D_{\sigma,\sigma} - \mu(\sigma)|$ versus word length $|\sigma|$ and $q$. See text for further detail.}
 \label{fig-deco02a}\end{figure}

How does then the classical limit arise in this context? In Figure \ref{fig-deco01} the {\em diagonal} matrix elements $D_{\sigma,\sigma}$ are plotted versus $q$, and compared with the classical probabilities $\mu(\sigma)$. Since the mass $M$ is proportional to ${\cal N} = 2^q$, increasing $q$ by one unit means doubling the mass of the particle. Convergence to the classical values is observed as well as the progressive quenching of off-diagonal matrix elements. In fact, in Figure \ref{fig-deco02a}, left panel, it is seen that when $q$ increases the decay of the sum of the absolute values of the off-diagonal matrix elements $D_{\sigma,\theta}$, $\sigma \neq \theta$, goes in pair with convergence to the classical entropy $S_{cl}$ of the {\em diagonal} entropy $S_{diag} = - \sum_\sigma D_{\sigma,\sigma} \log D_{\sigma,\sigma}$ and of the AF entropy $S_{AF}$ in \eqa{proj3}. The fitting line in the figure reveals that these phenomena scale as a power law with the mass $M$ (proportional to $2^q$), with exponent between minus one and minus one half.

The right panel of Figure \ref{fig-deco02a} reveals in even clearer detail the logarithmic catastrophe alluded before. The symbolic distance, $d_{\mbox{\tiny symb}} = \sum_\sigma |D_{\sigma,\sigma} - \mu(\sigma)|$, at fixed word length $|\sigma|$, is plotted when both $|\sigma|$ and $q$ are varied. One observes that classical word probabilities of fixed length $|\sigma|$ are recovered when increasing the mass of the particle ({\em i.e.} $q$), while quantum interference destroys this convergence when, at fixed mass, longer words are considered. This is the well-known non-commutativity of the classical limit and the long-time limit \cite{berry1,berry2}. Yet, the graph shows that quantum effects spoil the {\em correspondence} at a word length that is {\em logarithmic} in the mass of the particle. One so recovers from the entropic point of view the results obtained via algorithmic complexity \cite{physd}.

As mentioned, decoherence has been proposed as a way to overcome this {\em impasse}. It is the purpose of the remaining part of this paper to study decoherence in a controlled setting, with the aid of a model system.

\section{Decoherence and the multiparticle Arnol'd Cat}
\label{sec-multicat}

Decoherence, that is, the loss of quantum coherence of a system, is today principally studied because it is a major obstacle in the performance of a quantum computer \cite{strini}. In the nineties of the XX century was invoked for the explanation of the puzzling behavior of quantum Schr\"odinger cats \cite{haro1} and the emergence of classical properties in quantum mechanics \cite{joos}, \cite{ian1,halli,ian2,zurpaz1,zurpaz2}. Actually, as early as 1984 Guarneri \cite{italo1} proved that a random perturbation of the motion of a kicked rotor prevents quantum localization to set in. This induces a diffusive quantum motion, whose diffusion coefficient $D$ was computed by Ott {\em et al.} \cite{ott}. In the same perspective, Dittrich and Graham \cite{tomas}, Kolovsky \cite{andrey1} and Sundaram et al. \cite{arje1} considered systems (both classical and quantum) randomly perturbed or coupled to an environment: the resulting motion is also diffusive. The quantum dynamics of these system was shown to be similar to the classical when $\sqrt {D/ \lambda}$ is larger than $\hbar$ times a dimensional constant ($\lambda$ is the Lyapunov exponent). The heuristic explanation of this inequality stems from the observation that since diffusion {\em smears} classical phase space to the finest discernable resolution of $\sqrt {D/ \lambda}$, when this latter is larger than $\hbar$, the quantum smeariness, the two dynamics appear similar. To quote a famous musical,
{\em The way she sees and the way he looks is a perfect match}: the reader is asked to interpret the analogy. We shall return to this point in the conclusions.


Let us now describe a system that I have introduced to model {\em exactly} the interaction of a system with the environment, avoiding recourse to master equations, Lindblad operators, or the like, with the goal of having a fully Hamiltonian system \cite{gattosib}. The model draws upon the original view of decoherence by Joos and Zeh \cite{joos,teta1} as due to collisions of particles.

Consider adding $I$ smaller particles of mass $m$ in the ring that confines the particle of mass $M$, freely rotating except for collisions with the large particle.
Let $x_i$ and $y_i$, $i=1,\ldots,I$ the coordinates and momenta of these particles, respectively. The Hamiltonian of the classical system is therefore
\begin{equation}
 H = H_{cat}(X,Y,t) + \sum_{i=1}^I \frac{y_i^2}{2m}  + \kappa \sum_{i=1}^I
  V(x_i-X),
 \label{eq-ham10}
\end{equation}
where $H_{cat}$ has been introduced in \eqa{eq-ham1}, $\kappa$ is a coupling constant and the potential function $V(\cdot)$ will be described momentarily.

Let us first define the quantum kinematics.
In addition to the $L=1$ periodicity in the variables $x_i$, we also impose periodicity $\frac{mL}{T}$ in the momenta $y_i$. Therefore, performing a similar analysis with the one presented above, we obtain
\begin{equation}
   \frac{m L^2}{T} = { \nu} h,
  \label{eq-ham5a}
\end{equation}
where $\nu$ is the dimension of the Hilbert space of a single small particle, an integer that we also choose to be a power of two: $\nu=2^r$. 
Therefore, the manyparticle wave--function reads
\begin{equation}
  \Psi (X,x_1,\ldots,x_I) = {\cal N}^{-1/2} \nu^{-I/2} \sum_{k_0=0}^{{\cal N}-1}
   \sum_{k_1=0}^{\nu-1} \cdots
   \sum_{k_I=0}^{\nu-1}
   c_{k_0,k_1,\ldots,k_I} e^{2 \pi i(k_0 X + \sum_i k_i x_i)}.
  \label{eq-ham5m}
\end{equation}
In the same way as the position $X$ is restricted to a lattice, so are the $x_i$'s. We further require that small particles' positions $x_i$, $i=1,\ldots,I$, be a subset of those of the large one:
 \begin{equation}
  x_i^{j_i} = j_i \frac{1}{\nu} + s_i \frac{1}{{\cal N}}, \;\; j_i =0,\ldots,\nu-1.
  \label{eq-lat1}
\end{equation}
where $s_i$ an integer ranging from zero to ${\cal N}/\nu -1$.
Introducing the multi-index ${ j} = {j_0,j_1,\ldots,j_I}$, in which the label $0$ refers to the large particle and $i>0$ to the $i$-th small one, we let $|{j}>$ denote the delta function at the particles' position given in equations (\ref{eq-qbigj}), (\ref{eq-lat1}).  A different representation for the wave function is therefore
\begin{equation}
  \Psi  = 
   \sum_{j_0=0}^{{\cal N}-1}
   \sum_{j_1=0}^{\nu-1} \cdots
   \sum_{j_I=0}^{\nu-1}
   d_{j_0,j_1,\ldots,j_I}  |j_0,j_1,\ldots,j_I >.
  \label{eq-psi2}
\end{equation}
The Hilbert space of the system ${\cal H}$ is the tensor product of the Hilbert spaces of the single particles, and therefore it is isometric to a power of $\mathbf C$:
\begin{equation}
 {\cal H} = {\cal H}_{big} \otimes {\cal H}_{small}^I = {\mathbf C}^{{\cal N} I^\nu}.
  \label{eq-gatto1b}
 \end{equation}

The potential $V$ must represent a short--range interaction occurring only when the large particle and a small one occupy the same position. In the representation \eqa{eq-psi2} its matrix elements are therefore
\begin{equation}
 < { j'} | V { j}> = V_{j'_0,j'_1,\ldots,j'_I;j_0,j_1,\ldots,j_I} =
    \prod_{i=0}^I \delta_{j'_i,j_i}
   \sum_{i=1}^I \delta_{j_0, \frac{\cal N}{\nu} j_i + s_i}.
  \label{eq-scat3}
\end{equation}


\section{The Von Neumann Entropy}
\label{sec-von}

\begin{figure}
  \begin{center}
 \begin{tabular}{lll}
      \resizebox{40mm}{!}{\includegraphics[height=60mm, angle=-90]{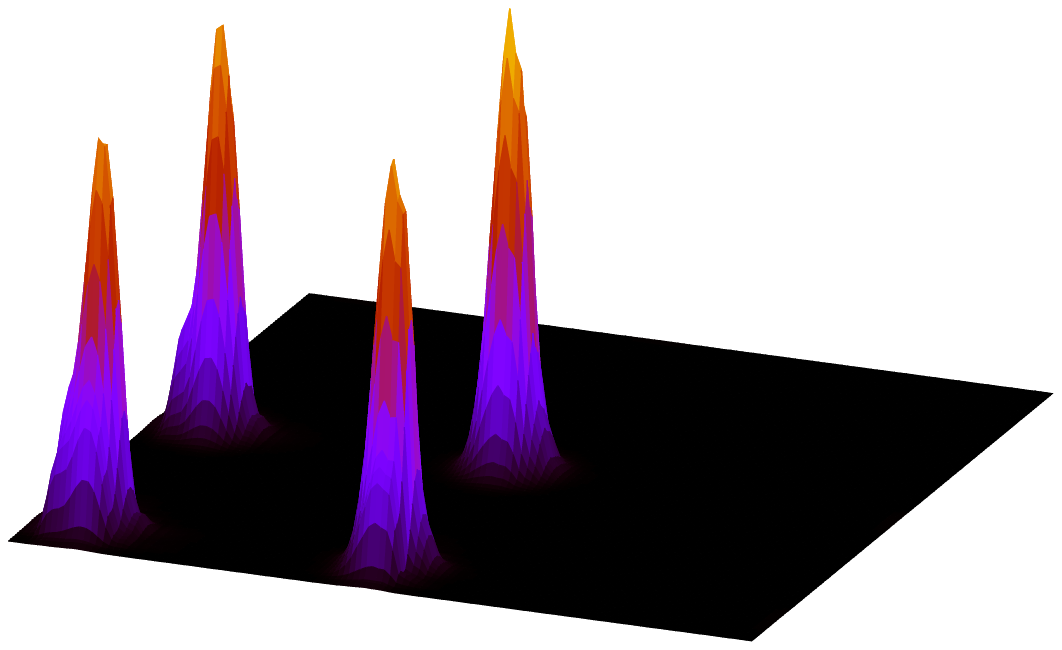}}   & \resizebox{40mm}{!}{\includegraphics[height=60mm, angle=-90]{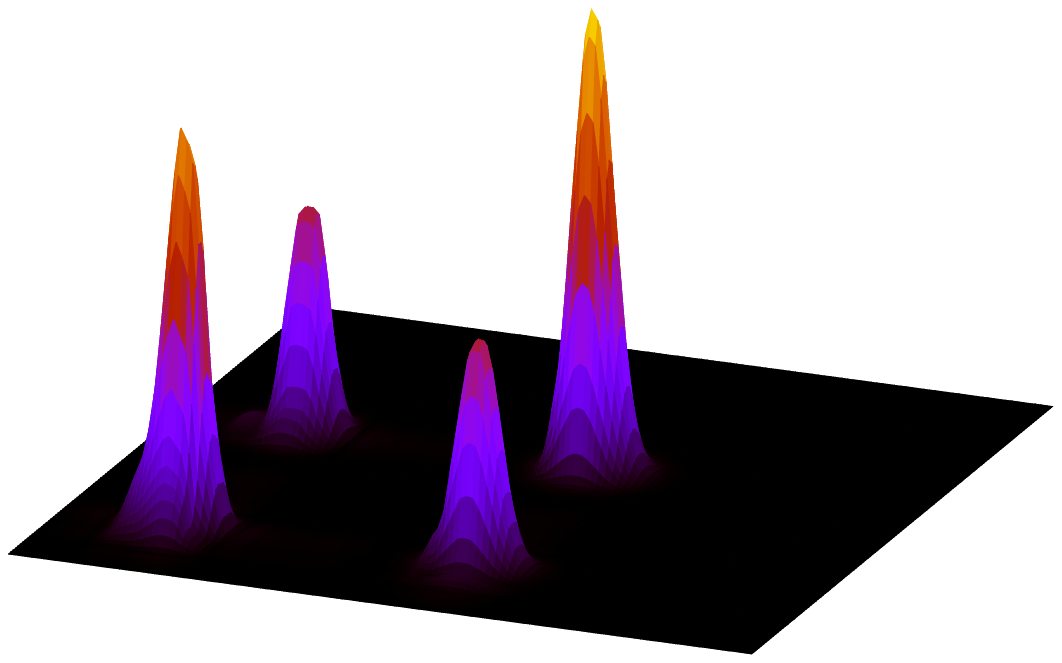}} &
       \resizebox{40mm}{!}{\includegraphics[height=60mm, angle=-90]{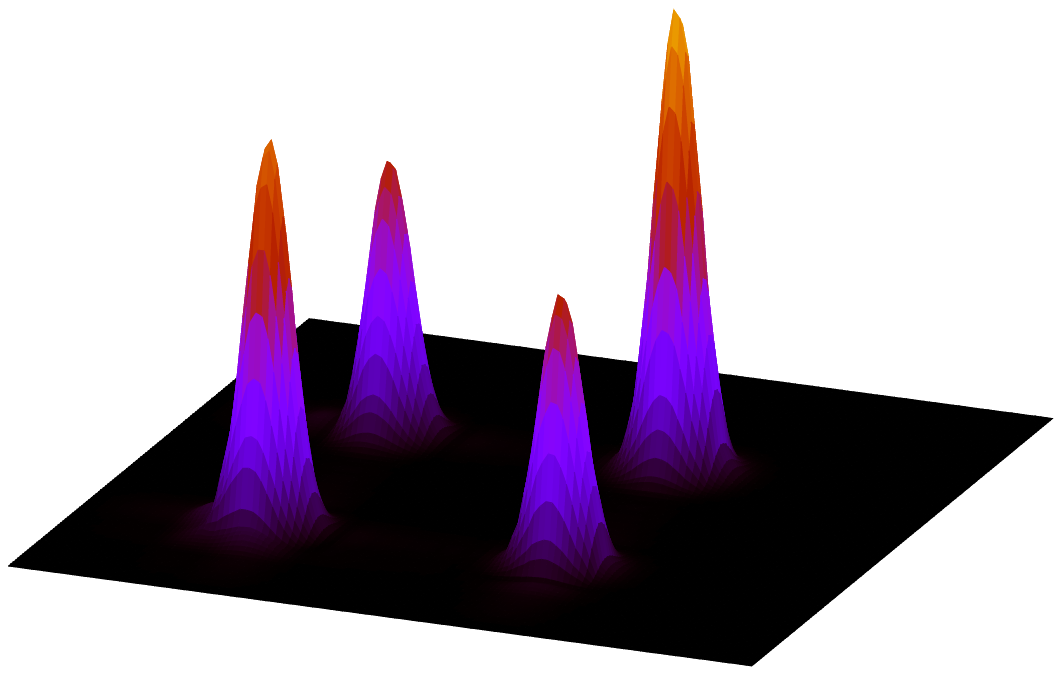}}    \\ \resizebox{40mm}{!}{\includegraphics[height=60mm, angle=-90]{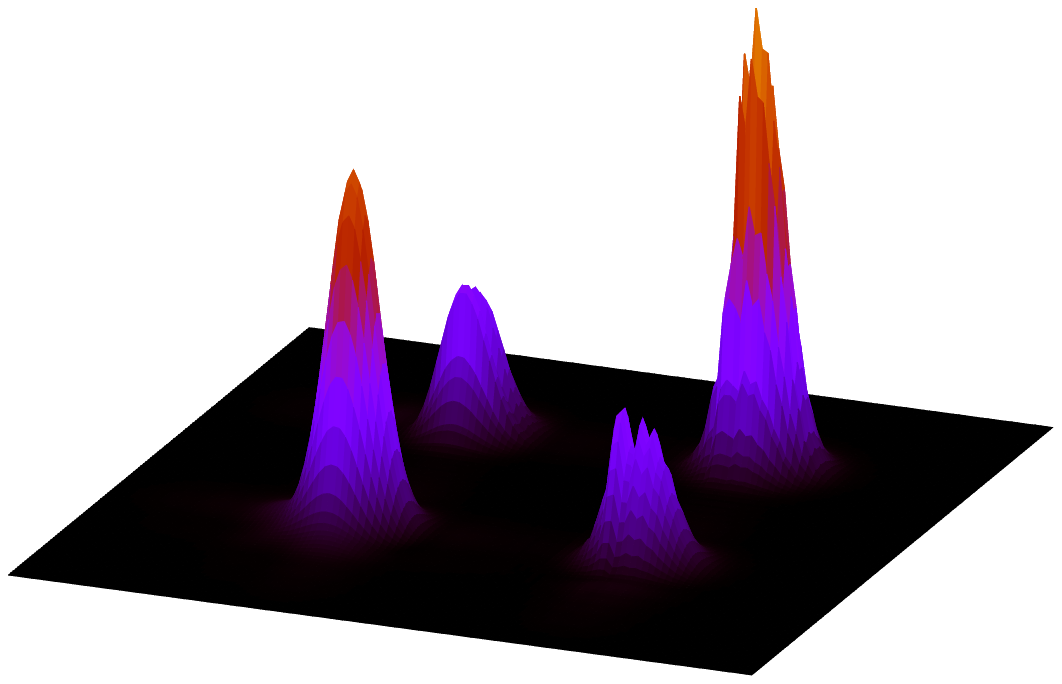}} &
       \resizebox{40mm}{!}{\includegraphics[height=60mm, angle=-90]{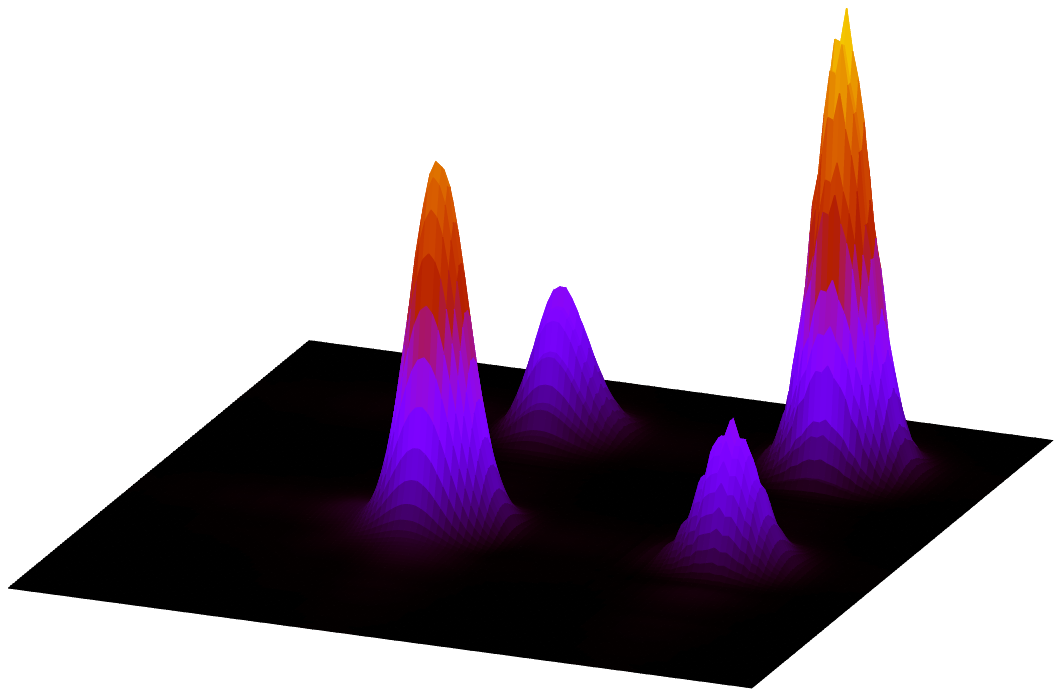}}   & \resizebox{40mm}{!}{\includegraphics[height=60mm, angle=-90]{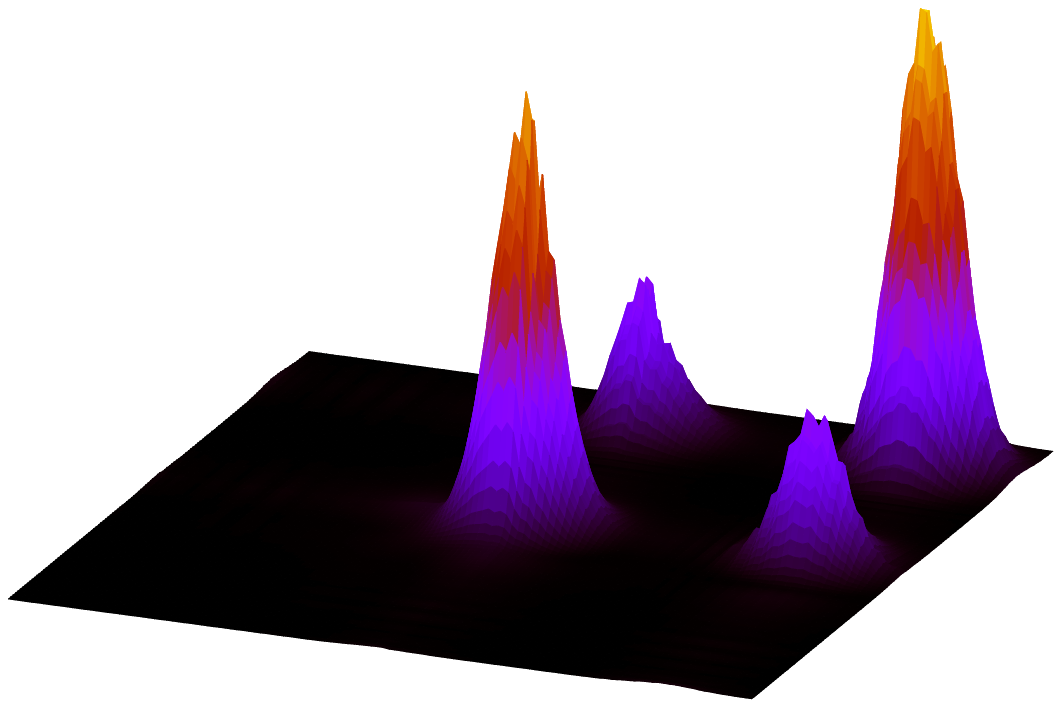}}
  \\
    \end{tabular}
  \end{center}
  \caption{Absolute value of the reduced density matrix, in the position representation, of a system with $q=7$, $I=2$, $\nu=1$, $\kappa=50$. Time snapshot ordered from left to right, top to bottom. The cat particle has a positive velocity, so that peaks drift on the torus.}
 \label{fig-scat}\end{figure}

In a pair of papers I have investigated the dynamics of the quantum system introduced in the previous section \cite{gattosib,condmat}. Treating the large, zeroth particle as the system and the small particles as the environment, the reduced density matrix $\hat \rho$ is defined by tracing over the small particles degrees of freedom:
\beq
 \hat \rho_{j'_0,j_0} = \sum_{j_1,\ldots,j_I}  \rho_{j'_0,j_1,\ldots,j_I;j_0,j_1,\ldots,j_I}
 \label{eq-trace}
 \end{equation}
Firstly, by choosing an initial density matrix as a pure state in a Schr\"odinger cat configuration \cite{haro1}, I showed that as time progresses, interaction with small particles wipes out the outdiagonal components of the reduced density matrix $\hat \rho$, obtained by tracing over the small particles degrees of freedom, exactly as in the experimental results. This dynamics is shown in the position representation in Figure \ref{fig-scat}. Here, $\alpha=0$, that is, there is no cat kick. These results show that the multiparticle cat is a {\em bona fide} model of decoherence.

I then considered the Von Neumann entropy, $- \mbox{Tr}  \; \hat \rho \log \hat \rho$. This entropy quantifies the amount of information in the eigenvalues of the reduced density matrix.
The analysis was first performed for the free colliding particles case, given by the Hamiltonian
(\ref{eq-ham2}), (\ref{eq-ham10}) with $\alpha=0$. The initial increase of the Von Neumann entropy was observed to be logarithmic. On the contrary, when $\alpha T^2/M=1$, as in the Arnol'd cat map, the initial increase of entropy is linear, with a slope that equals the Kolmogorov Sinai entropy {\em provided} the coupling constant $\kappa$ (Eq. \ref{eq-ham10}) takes a {\em specific} value, which scales according to the physical parameters of the system. Clearly, when $\kappa$ is null there is no decoherence and the Von Neumann entropy is null. For small values of $\kappa$ entropy increases linearly at a small rate before saturation, while above the aforementioned value the rate is larger than the classical dynamical entropy. This parameter dependence is in my view a {\em caveat} for the physical interpretation of similar results in different systems. For details, the interested reader is referred to \cite{condmat}.

\section{The AF Entropy of the multiparticle Arnol'd cat}
\label{sec-multiaf}

Even though interesting, results about the Von Neumann entropy lack the generality and theoretical significance provided by the Alicki-Fannes entropy. In particular, the former is bounded above by the logarithm of the dimension of the Hilbert space of the {\em reduced} system, in this case $\log {\cal N}$, {\em independently} of the dimension of the {\em full} Hilbert space. The latter, on the contrary is bound by  $2 \log {\cal N} + 2 I \log(\nu)$, which scales {\em linearly} with the number of particles $I$ and therefore {\em might} provide a physically reasonable classical limit.

Moreover, the decoherence matrix introduced in Sect. \ref{sec-entropies} permits to compare analytically, {\em i.e.} one by one the classical probabilities $\mu(P_\sigma)$ with the diagonal elements $D_{\sigma,\sigma}$ and to gauge the importance of non diagonal matrix elements, which encode quantum coherence. Observe that the projectors $P_l$ in \eqa{proj1} can be generalized to act on the coordinate of the large particle only, labeled by the index zero:

 \begin{equation}
  P_l = \sum_{j_0=l Q}^{(1+l)Q-1}
  \sum_{j_1=0}^{\nu-1} \cdots  \sum_{j_I=0}^{\nu-1}
  |j_0,j_1,\ldots,j_I ><j_I,\ldots,j_1,j_0|
  , \;\;\; l = 0,\ldots,2^{p}-1.
  \label{projbig}
 \end{equation}

In so doing, we are observing the symbolic dynamics of the large particle only, in the presence of the external environment provided by the other, smaller, particles.

 \subsection{The decohering effect of the perturbation}
\begin{figure}
   \begin{center}
 \begin{tabular}{cc}
      \resizebox{80mm}{!}{\includegraphics[height=60mm, angle=-90]{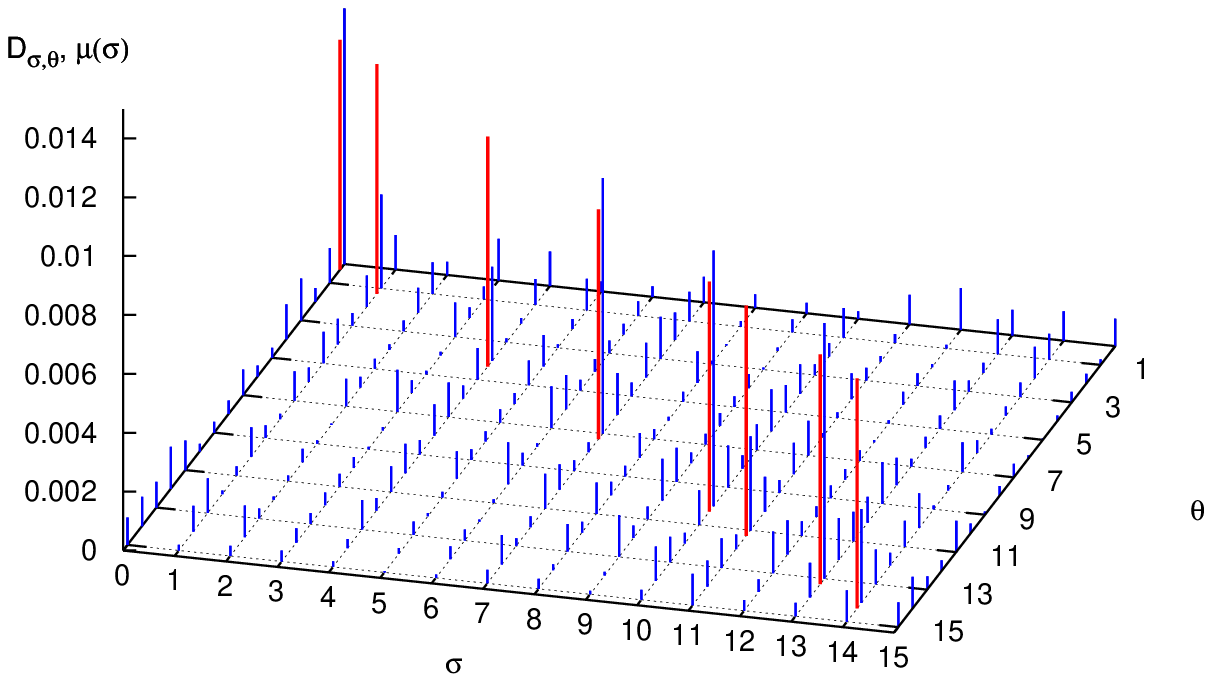}}   & \resizebox{80mm}{!}{\includegraphics[height=60mm, angle=-90]{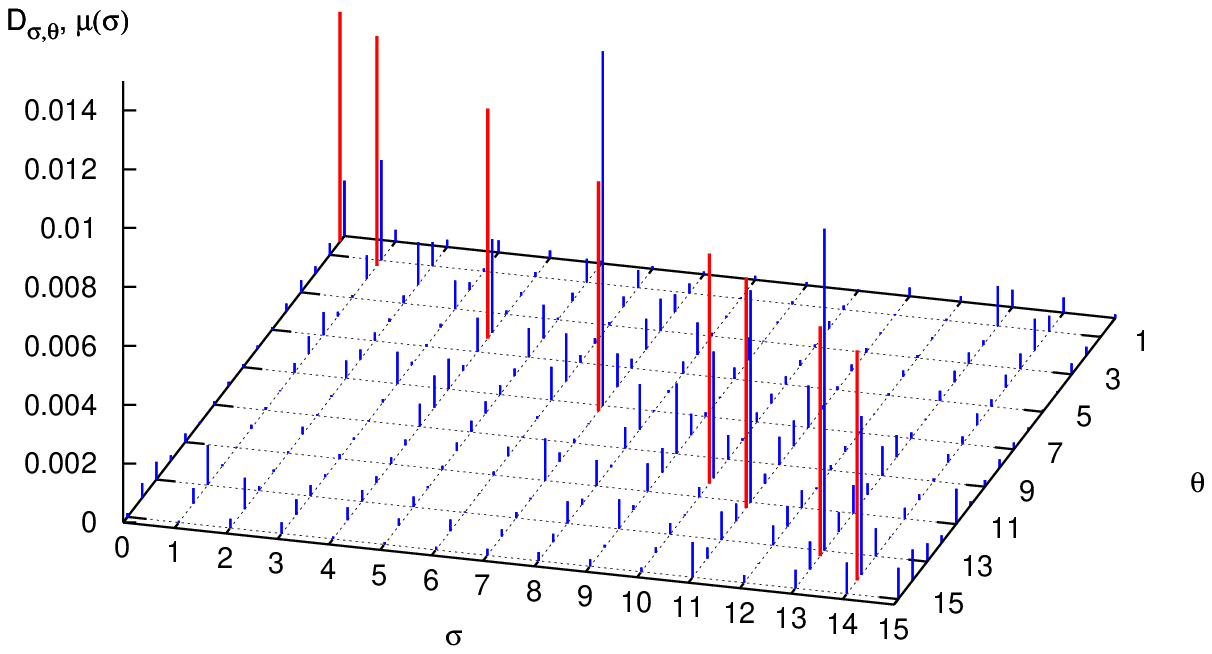}}
  \\
    \end{tabular}
  \end{center}

  \caption{Absolute value of the decoherence matrix elements $D_{\sigma,\theta}$ (blue bars) and classical word probabilities $\mu(\sigma)$ (red bars, plotted on the diagonal for comparison) when $q=4$, $r=1$, $I=2$ and $p=2$. In the left and right panels $\kappa=0$ and $\kappa=10$, respectively. See text for further detail.}
 \label{fig-sun-13a}\end{figure}

\begin{figure}
  \begin{center}
 \begin{tabular}{cc}
      \resizebox{80mm}{!}{\includegraphics[height=60mm, angle=-90]{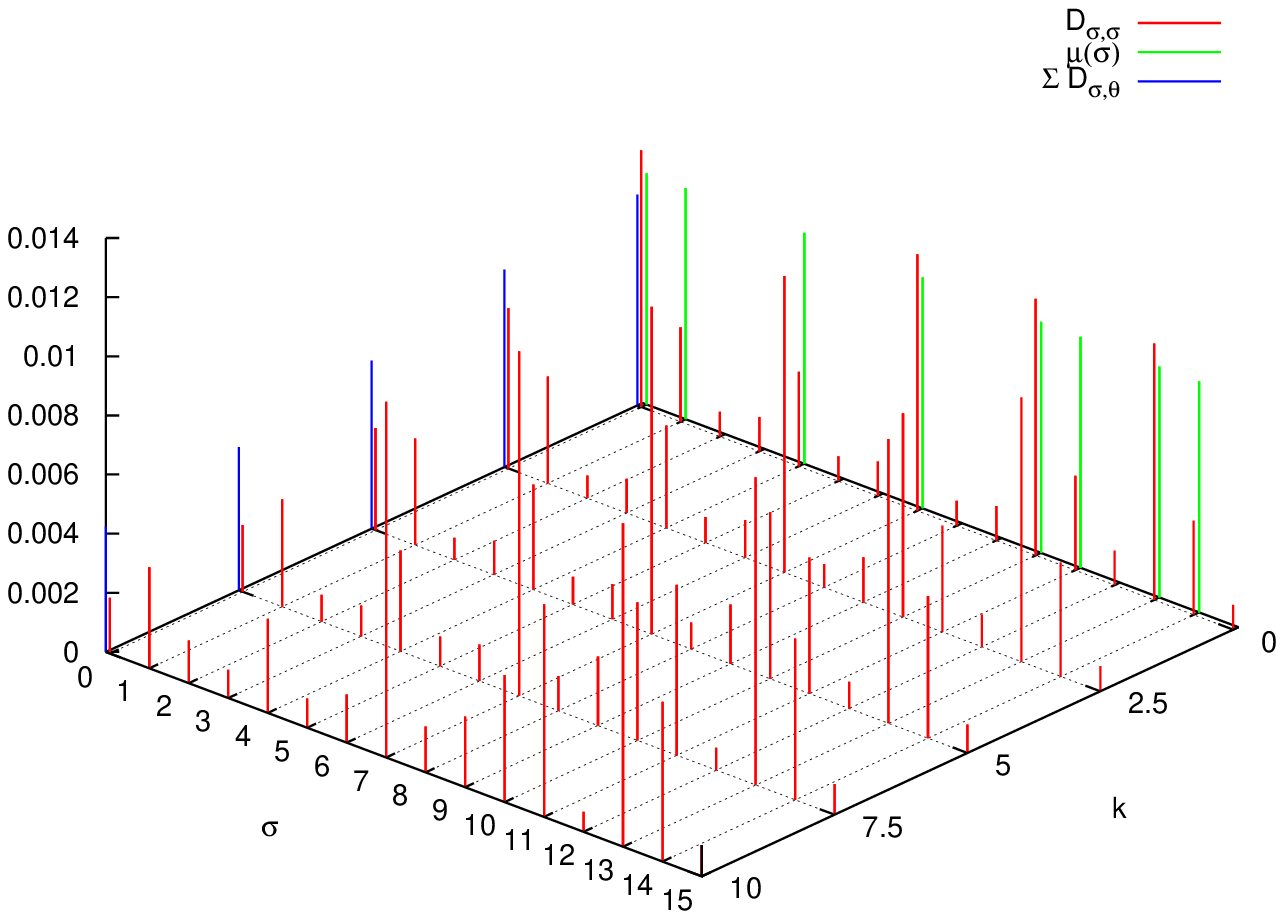}}   & \resizebox{80mm}{!}{\includegraphics[height=60mm, angle=-90]{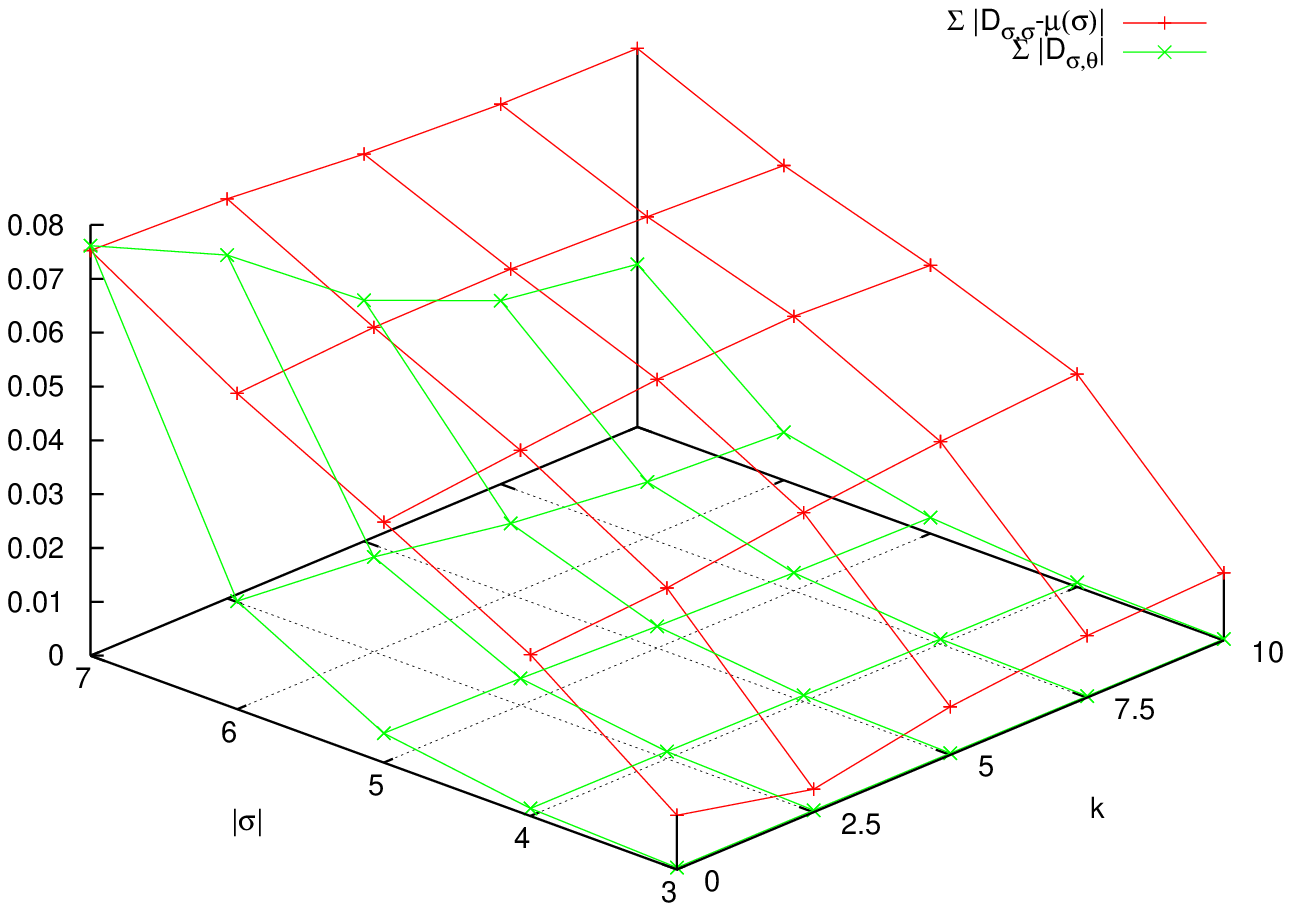}}
  \\
    \end{tabular}
  \end{center}
  \caption{Left panel: Parameters as in Fig. \ref{fig-sun-13a}. Diagonal matrix elements $D_{\sigma,\sigma}$ (red bars) versus $\kappa$, classical word probabilities $\mu(\sigma)$ (green bars), and sum of the absolute value of outdiagonal matrix elements $D_{\sigma,\theta}$ (blue bars, rescaled by a factor to improve readability of the graph).  Right panel: sum of the absolute differences between diagonal matrix elements and classical probabilities (red), sum of the absolute values of off-diagonal matrix elements (green, rescaled by a factor).}
 \label{fig-sun13b}\end{figure}

Firstly, let us consider a system composed of a large particle, $q=4$, coupled to two small particles, $r=1$ (mass ratio 1:8) and let us vary the coupling constant $\kappa$. 
In Figure \ref{fig-sun-13a} we plot as before the decoherence matrix entries for words of length 5, beginning and ending with the symbol 0. In the left panel $\kappa=0$ (that is, small particles do not interact with the large one), while in the right panel $\kappa=10$. Non diagonal matrix elements are clearly smaller in the second case, so that diagonal values become closer to the  eigenvalues of the decoherence matrix. Nonetheless, we do not expect them to approach {\em in the limit of very large $\kappa$} the classical word probabilities, displayed as red bars in both panels. This is evident in Figure \ref{fig-sun13b} where the diagonal matrix elements $D_{\sigma,\sigma}$ are plotted versus $\kappa$ and compared to $\mu(\sigma)$.  The symbolic distance $d_{\mbox{\tiny symb}}= \sum_\sigma |D_{\sigma,\sigma}-\mu(\sigma)|$ quantifies the agreement: in the right panel of Figure \ref{fig-sun13b} it is seen that $d_{\mbox{\tiny symb}}$ does not significantly diminishes when increasing $\kappa$, differently from the
sum of the absolute values of the off-diagonal matrix elements, $\sum_{\sigma \neq \theta} |D_{\sigma,\theta}|$. 

\begin{figure}
  \begin{center}
 \begin{tabular}{cc}
 \resizebox{80mm}{!}{\includegraphics[height=60mm, angle=-90]{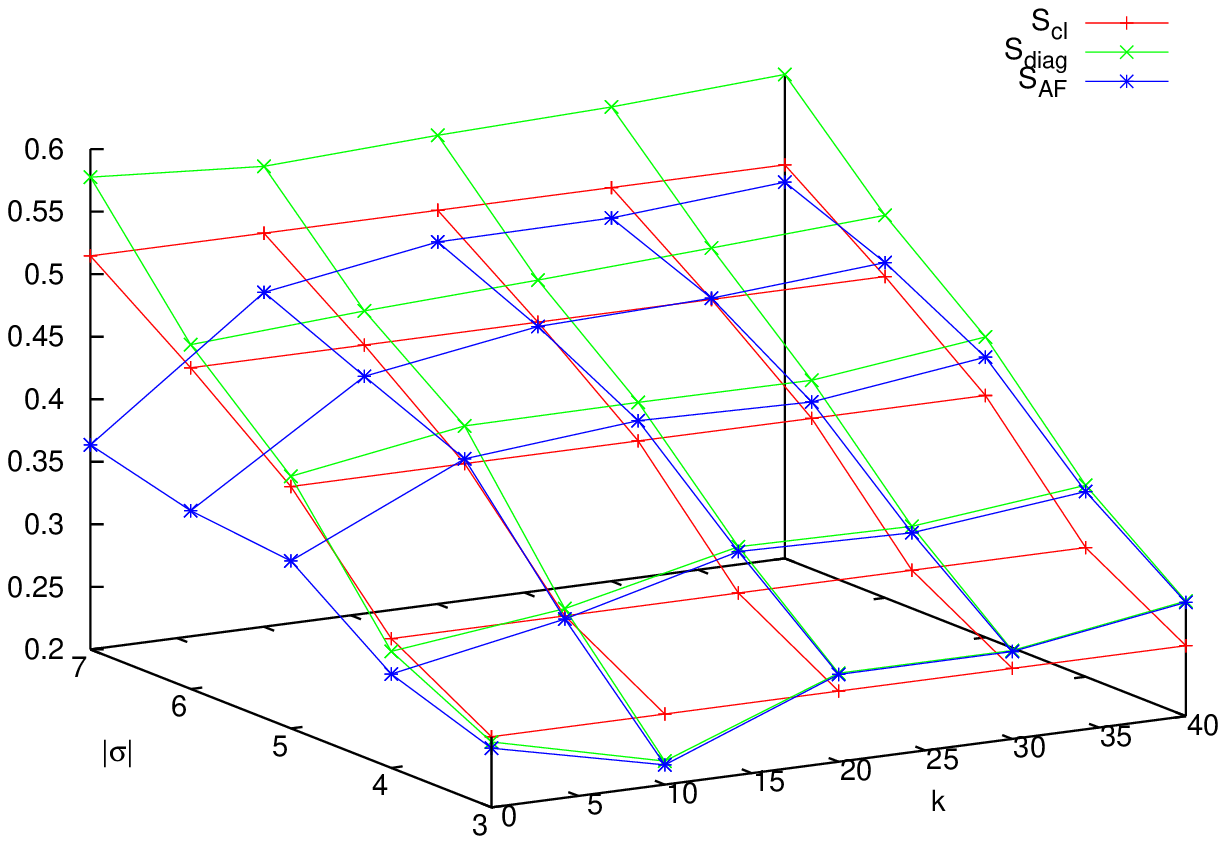}}   & \resizebox{80mm}{!}{\includegraphics[height=60mm, angle=-90]{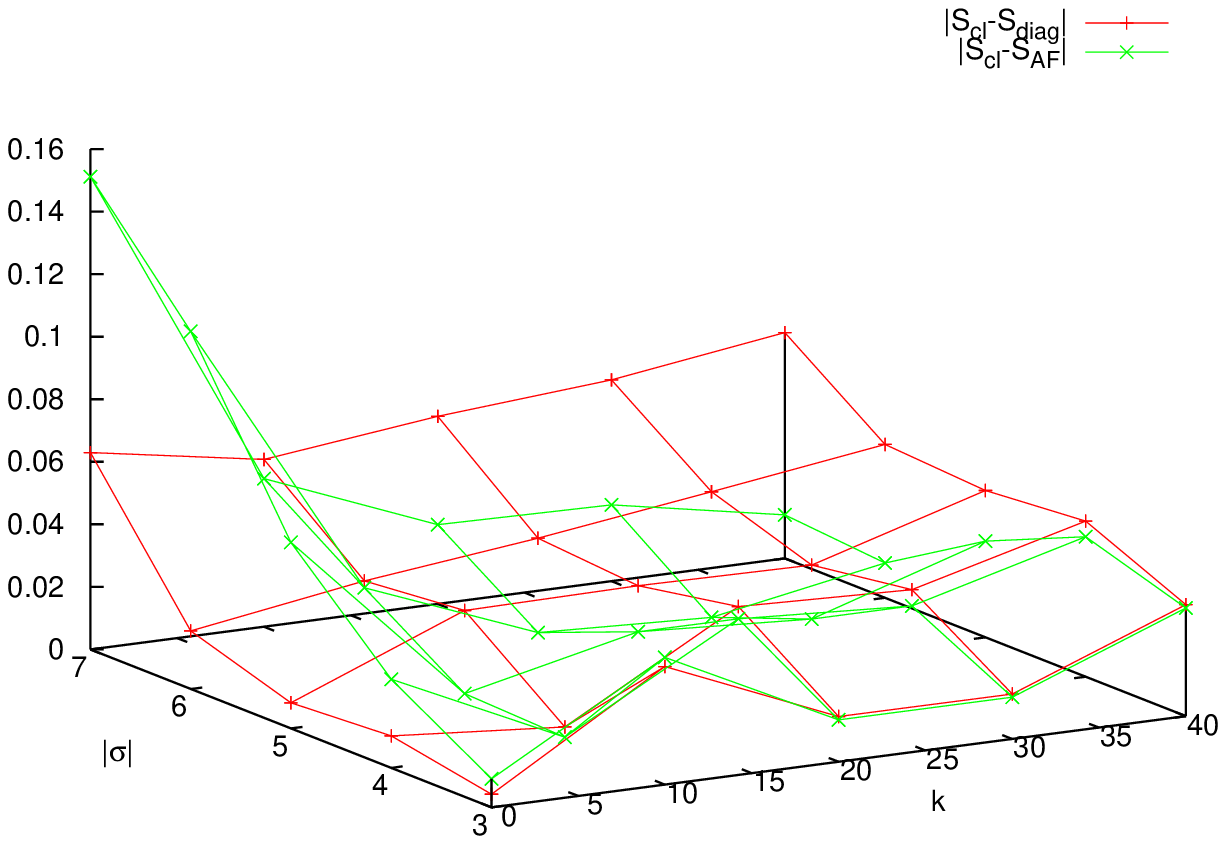}}
  \\
    \end{tabular}

  \end{center}
  \caption{Parameters as in Fig. \ref{fig-sun-13a}, wider range of the coupling constant $\kappa \in [0,40]$. Left panel: classical entropy $S_{cl}$ (red curves), { diagonal} entropy $S_{diag}$ (green curves), AF entropy $S_{AF}$(blue curves). Right panel: absolute difference $|S_{cl}-S_{diag}|$ (red curves) and $|S_{cl}-S_{AF}|$ (green curves).}
\label{fig-deco03}
\end{figure}

I believe that these results are general, albeit obtained in a restricted range of parameters.
Firstly, increasing $\kappa$ reduces the off-diagonal matrix elements and therefore makes the system less coherent, {\em i.e.} more classical. Secondly, observe that the sums $\sum_\sigma D_{\sigma,\sigma}$ and $\sum_\sigma \mu(\sigma)$ over words of fixed length $|\sigma|$ are constant and equal to the probabilities, quantum and classical, of the sector $\sigma_0,\sigma_{n-1}$, ($0,0$ in the figures) so that changing the value of $\kappa$ results in a transfer of probability among all words of length $n=|\sigma|$ while keeping their sum approximately constant (because transfer of probability among different blocks $\sigma_0, \sigma_{n-1}$ is small). Therefore, the fact that increasing $\kappa$ at fixed $|\sigma|$ does not reduce the symbolic distance means that small particles are rendering the {\em observation} of the large one more classical, but at the same time are perturbing it too much.
Finally, the increase of the symbolic distance with $|\sigma|$ (an evident proxy for time) at fixed $\kappa$ shows again the rapidly emerging difference between classical and quantum dynamics. Figure \ref{fig-deco03} confirms this conclusion: we see that in the presence of decoherence (that is, low off-diagonal matrix elements) AF and diagonal entropies approach each other, but differ from the classical value.

\subsection{The limit of large mass with decoherence}

The previous results have shown that there exist two ways to induce decoherence, defined as quenching of the off-diagonal entries of the matrix $D$: increasing the mass of the particle (that is, the dimension of its Hilbert space), or coupling it to a larger, yet still purely quantum system. The former leads in the limit to the classical results, but is plagued by its very slow, logarithmic convergence, which renders its application unphysical. The second cannot provide by itself the hardly-sought classical limit, but it may feature an exponential increase in Hilbert space dimension, simply augmenting the number of small particles.
I made this observation in \cite{gattosib}, where I surmised that a combination of the two approaches might yield scaling relations (in the number of small particles, coupling, and mass of the large particle) by which complexity of the motion, as a linear increase of the Shannon Alicki-Fannes entropy $S(n)$, could be observed over physically significant time spans.

\begin{figure}
\includegraphics[height=80mm, angle=-90]{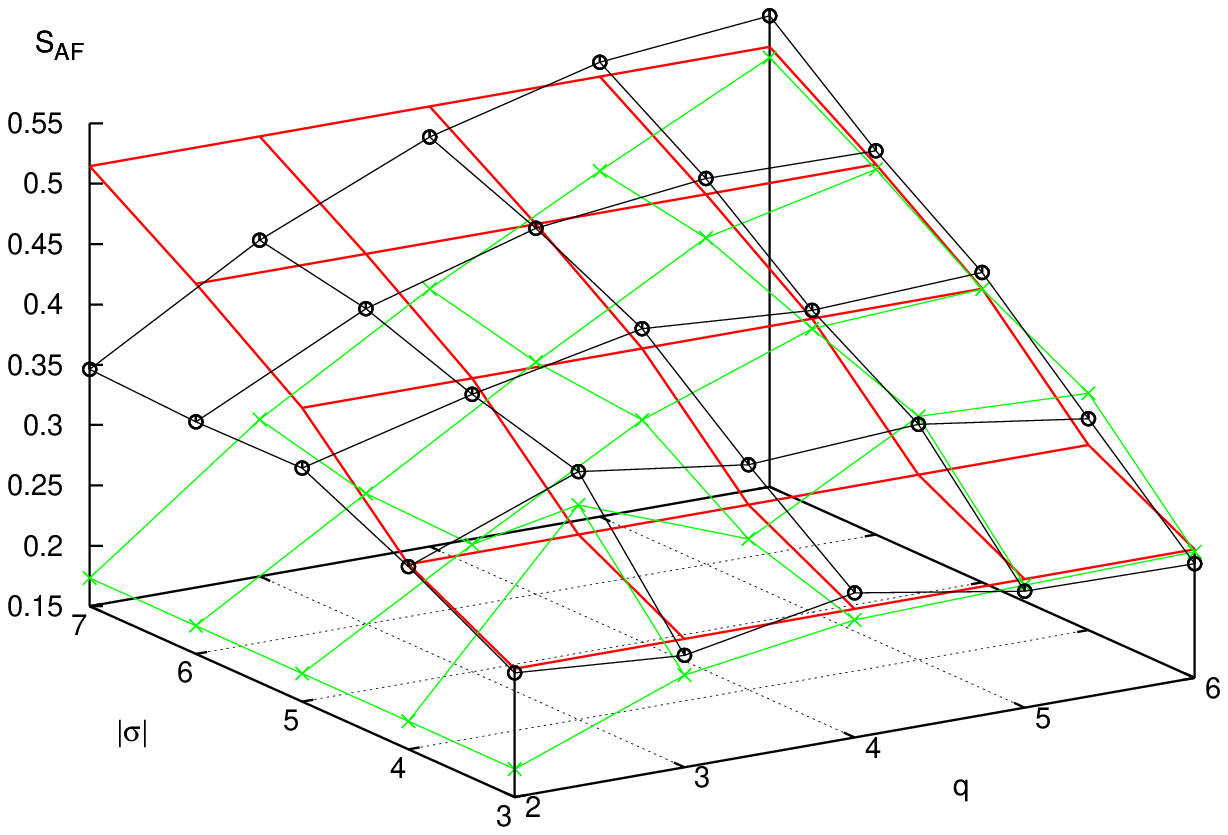}
  \caption{Classical entropy $S_{cl}$ (red curves) compared with AF entropy $S_{AF}$ of systems with variable $q$ and $I=0$ (green curves) and $I=2$, $r=1$ and $\kappa=.125 \times 2^q$ (black curves).
 }
\label{fig-deco023}\end{figure}

This investigation is subtle, both from the theoretical and the numerical viewpoint. In this paper we can add a few experiments in this direction. In Figure \ref{fig-deco023} we consider two families of systems, the former composed of a single particle, of increasing mass (different values of $q$), already studied in Section \ref{sec-afcat}; the second, composed again of such large particle, but coupled to two small particles, with coupling constant $\kappa$ that increases proportionally to the mass of the large one. We compute the AF entropies $S_{AF}$ of these systems and compare them with the classical entropy. We observe that, while both converge to the last quantity, the multiparticle case provides a better approximation for small values of $q$. Considering that the mass $M$ is proportional to the exponential of $q$, this is encouraging in the quest outlined above.

\begin{figure}
  \begin{center}
 \begin{tabular}{cc}
      \resizebox{80mm}{!}{\includegraphics[height=60mm, angle=-90]{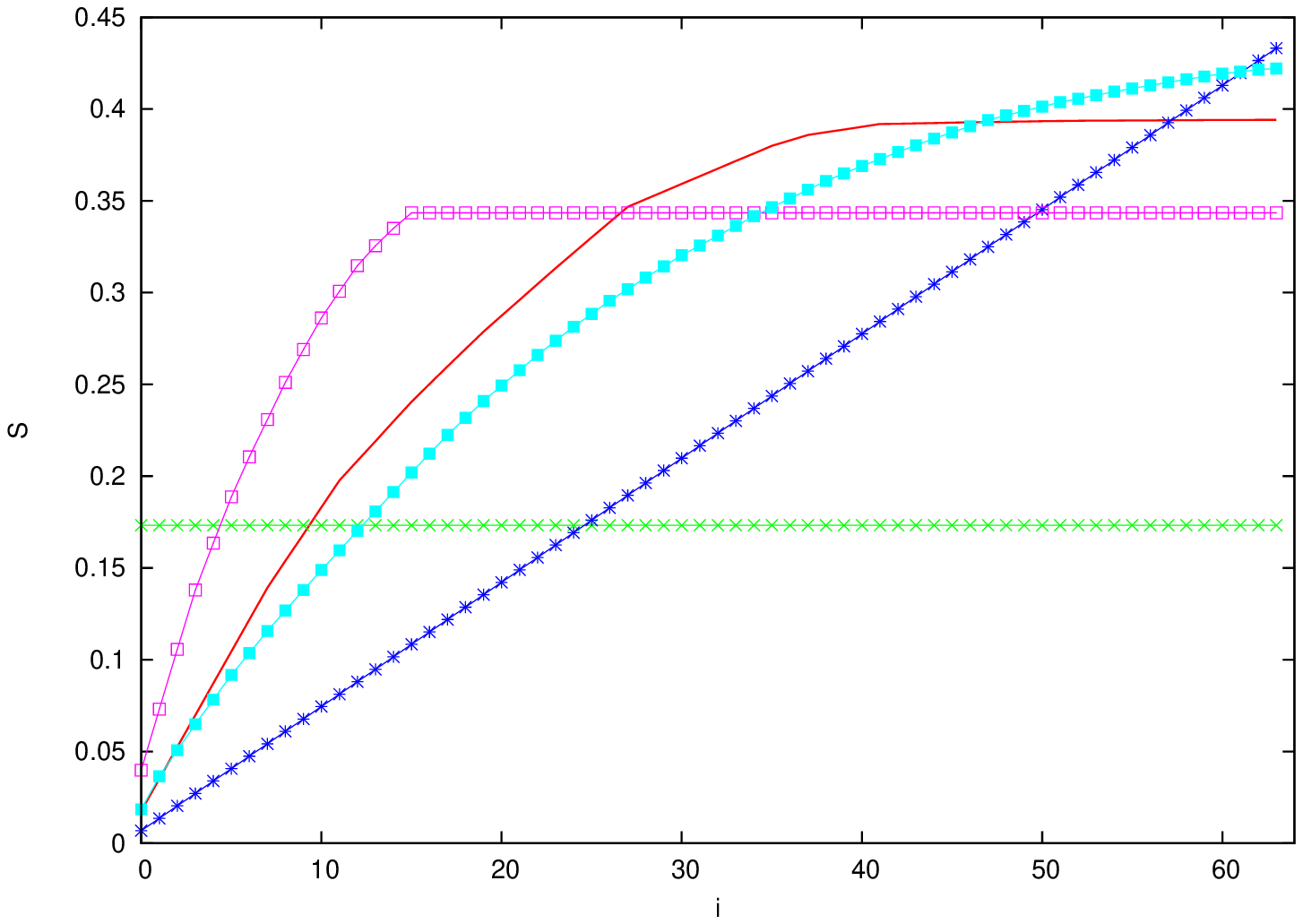}}   & \resizebox{80mm}{!}{\includegraphics[height=60mm, angle=-90]{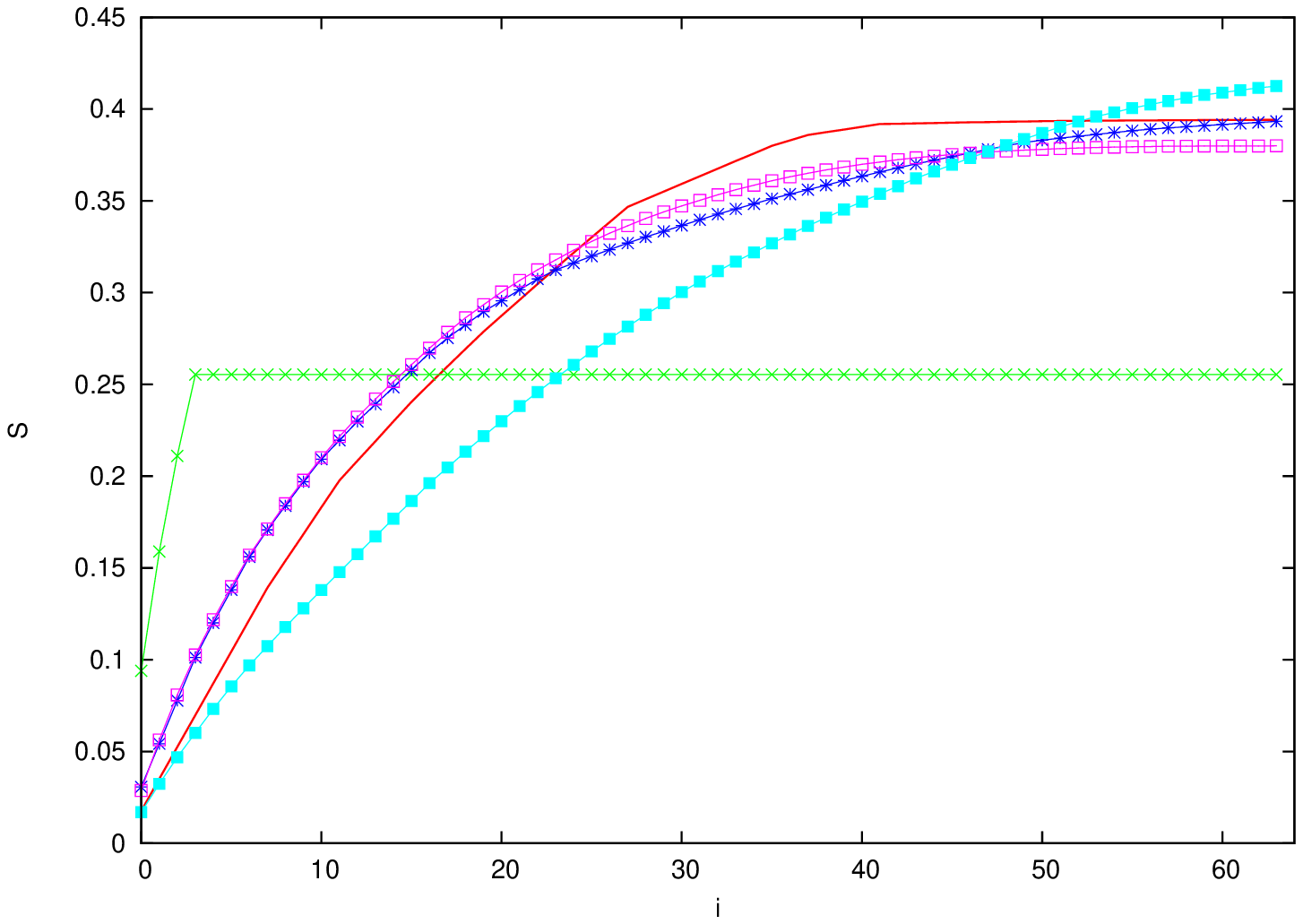}}
  \\ \resizebox{80mm}{!}{\includegraphics[height=60mm, angle=-90]{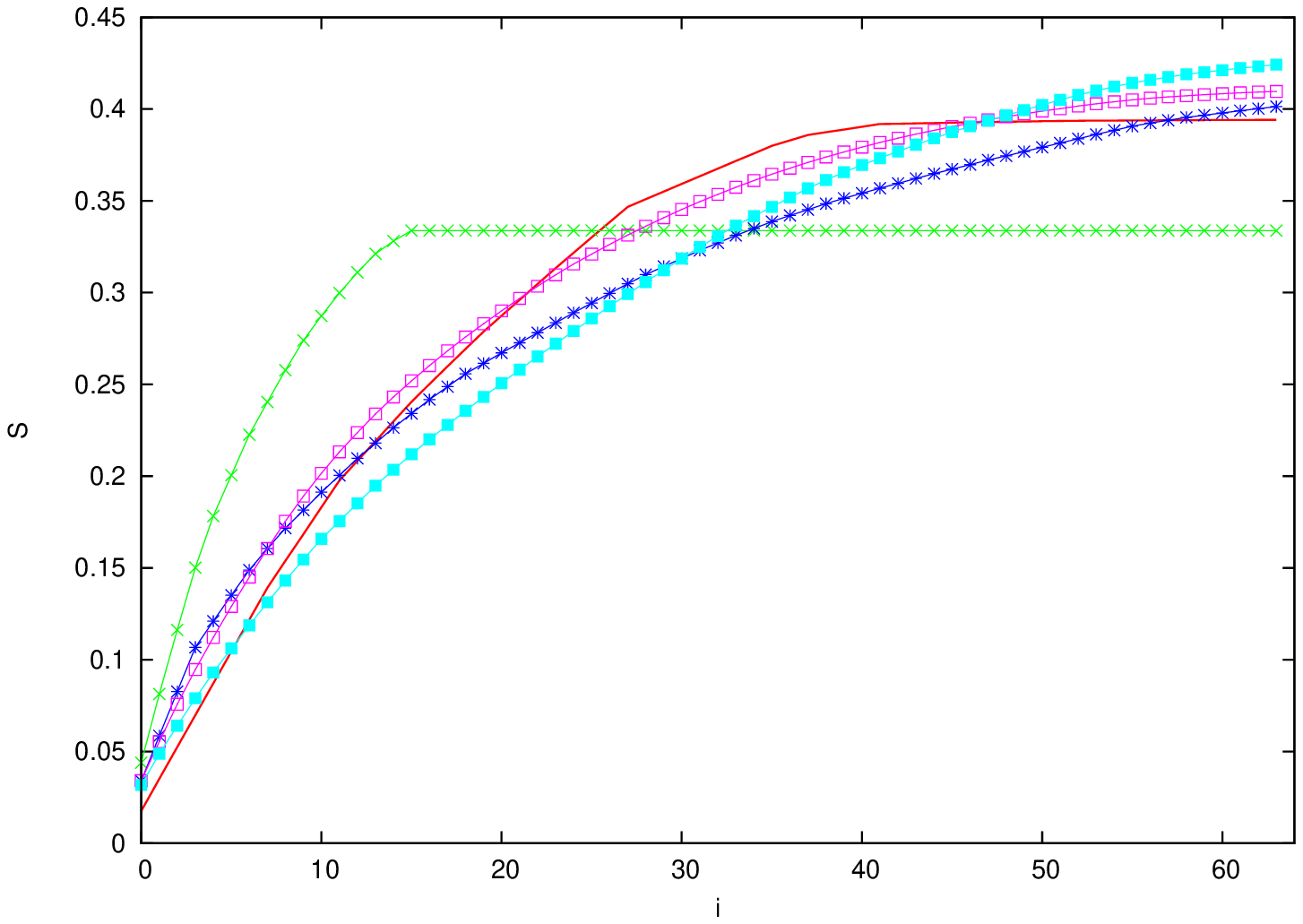}}   & \resizebox{80mm}{!}{\includegraphics[height=60mm, angle=-90]{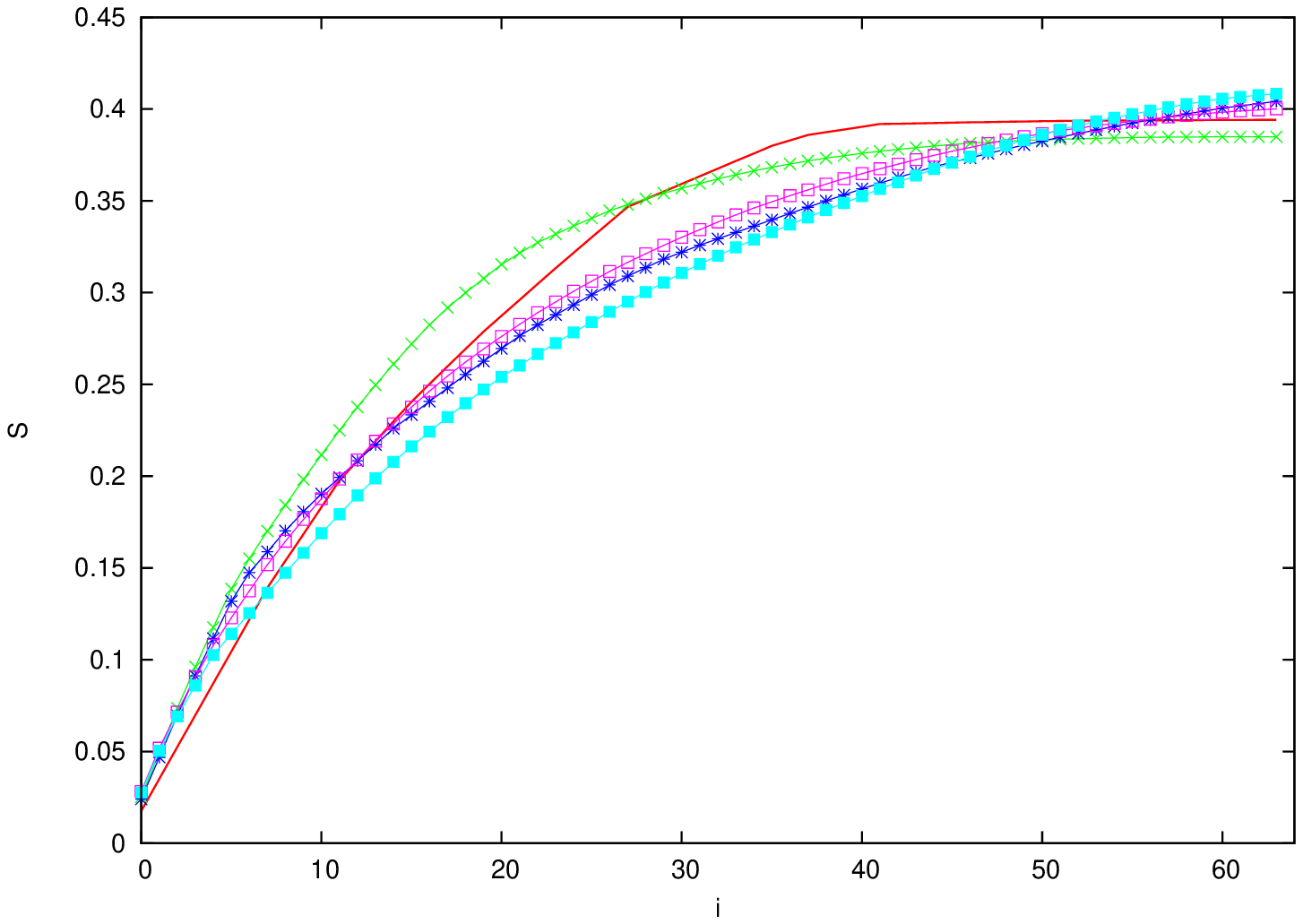}}
  \\ \resizebox{80mm}{!}{\includegraphics[height=60mm, angle=-90]{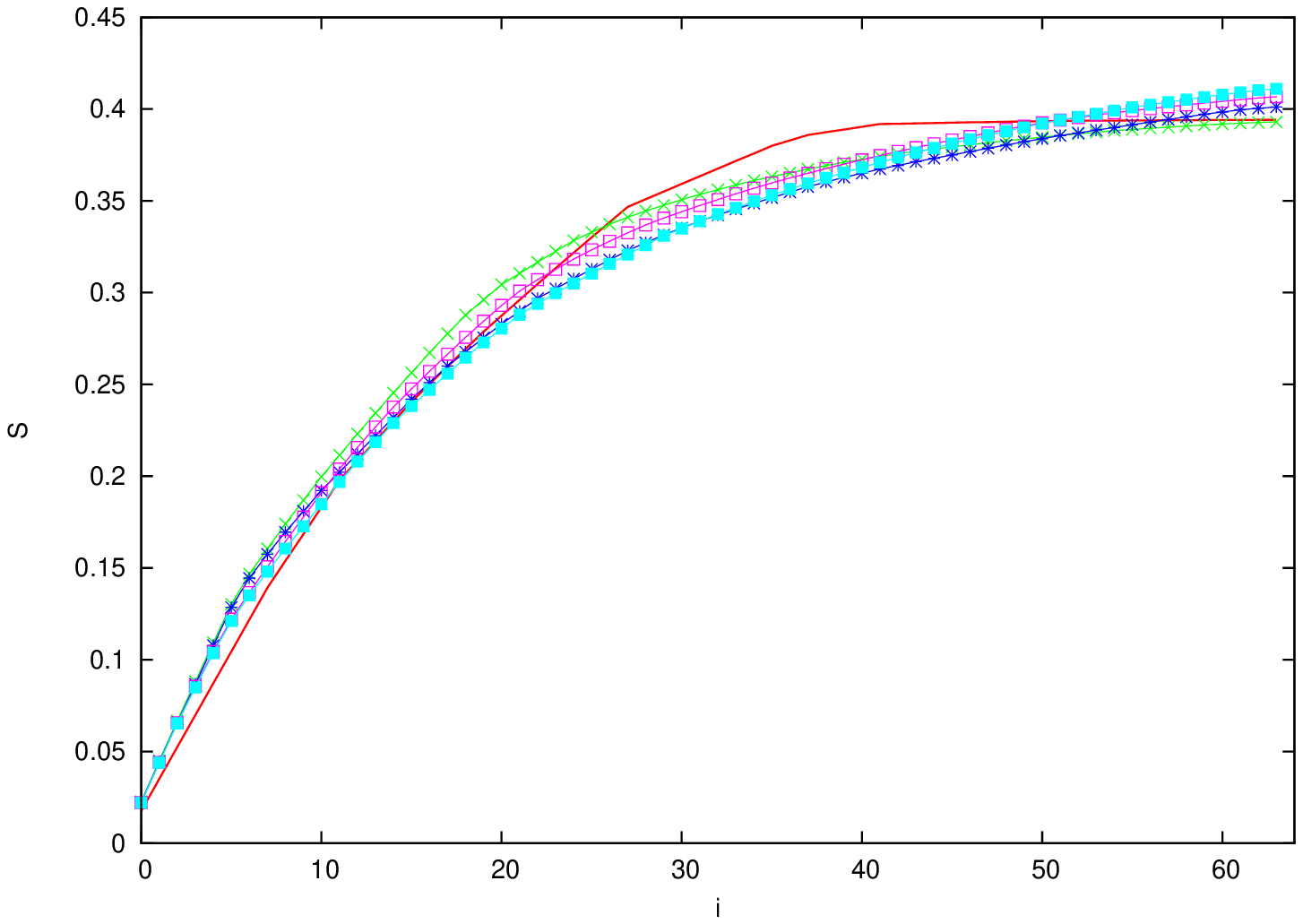}}   & \resizebox{80mm}{!}{\includegraphics[height=60mm, angle=-90]{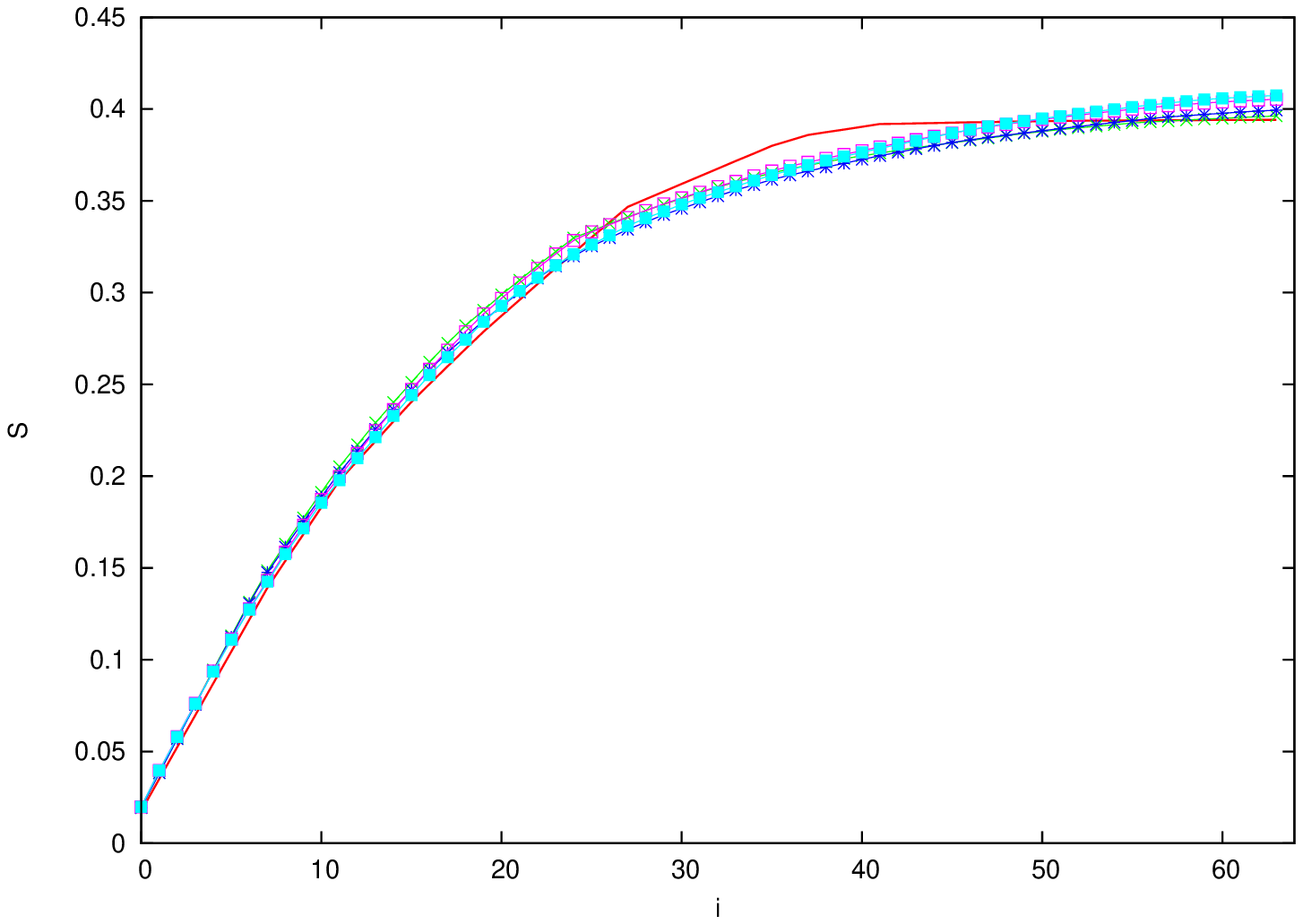}}
  \\
    \end{tabular}

  \end{center}
  \caption{Increasing partial entropies $S_i$, $|\sigma|=5$, of systems with variable $q$, from 2 to 7, left to right and top to bottom. The classical entropy $S_i$ is denoted by a red curve in all panels. In the case $I=0$ the green curve marks the AF partial entropy, and the blue curve the diagonal partial entropy. In the case $I=2$, $r=1$ and $\kappa=.125 \times 2^q$ the AF partial entropy is marked in magenta and the diagonal partial entropy in light blue.
  }
\label{fig-deco022}\end{figure}

To investigate this phenomenon, recall that,
as described above, for any finite value of $n=|\sigma|$ and for any finite sector $\sigma_0,\sigma_{n-1}$, ($0,0$ in the figures) the AF, diagonal, and classical entropies are the sum of $L^{n-1}$ contributions of the form
$s_j = -\zeta_j \log(\zeta_j)$, when $\zeta_j$ is an eigenvalue of the matrix $D$, or a diagonal entry of this matrix, or a classical word probability, respectively. We can therefore compare different entropies by first sorting the values $s_j$ in descending order, from the largest to the smallest, and then computing their integrated distribution $S_i = \sum_{j=0}^i s_j$. These distributions are plotted in Figure \ref{fig-deco022}, in the same case of Figure \ref{fig-deco023}. Different panels are related to increasing values of $q$, from 2 to 7. The horizontal segments in the plots of the AF partial entropies are due to the Hilbert space dimensionality bound, which imposes a maximum number of non-null eigenvalues of $D$. We can observe that the effect of the small particles is to free the AF entropy from this bound, assuring a faster convergence to the classical values.

\section{Conclusions}

The multiparticle Arnol'd cat is a model system particularly suited to investigate the nature, definition and relevance of quantum chaos. In my works I have followed the original suggestion of Chirikov and Ford that this problem should be considered within the framework of information theory. Dynamical entropies provide a major tool for this investigation and their formal aspects have been fully developed in the last years \cite{benabook}.
In this respect, the model described herein permits to free dynamics in a controlled, fully canonical way, from the bound induced by the finite dimensionality of the Hilbert space. This bound is due, in this particular model, by the finiteness of the mass of a particle, and therefore carries a deep physical significance.

In fact, I considered herein the classical limit as the limit of increasing mass of such particle, while keeping the value of the Planck constant fixed. Chaotic behavior has been defined, for a finite time span, as the production of information at an inferiorly bounded, positive rate. Consideration of the decoherence matrix from the consistent histories formalism has also permitted to compare word by word classical and quantum symbolic dynamics. It came as no surprise that, as I presented in earlier works, in the single particle case the length of agreement between the two, and consequently the length of the span of the chaotic freedom of the quantum system, grows only logarithmically with the mass of the particle.

As reviewed in this paper, coupling to an external system has been shown to provide quantum motions (via the Wigner function, or similar phase space tools) that {\em correspond} to a classical {\em noisy} evolution, within the finite phase-space resolution of this latter, for longer times. In a discrete lattice model, mimicking the quantum cat Hamiltonian with noise \cite{disc1} we have observed information production over long time scales. Returning now to many particles quantum systems, in \cite{gattosib} I put forward the hypothesis that a particular scaling of parameters: number of particles, their mass, and the observational time interval, could extend in a physically reasonable range the span of chaotic behavior. The results of this paper are admittedly not conclusive, but suggest that this endeavor could be feasible.
In this view, the (fully quantum) environment is to be considered as a source of {\em disorganized} information that, when input to a regular or chaotic quantum system, is {\em organized} by this latter in such a way to agree, within observational limits, to the classical description. This is in essence the program outlined in \cite{disc1}:
{\em For a long time, research in quantum chaos has looked for quantum characteristics related to classical chaotic motion. The fact that none of these could be properly called chaos led to the concept of pseudochaos and cast doubts on the very existence of chaos in nature. [..] one might try to reverse this approach and consider classical dynamics as an effective theory that, via truly chaotic deterministic dynamical systems, models a randomly perturbed quantum motion under observational coarse graining.}

\section{Acknowledgements}
Partially funded from grant PRIN 2017S35EHN {\em Regular and stochastic behavior in dynamical systems}.
Computations for this paper were performed on the Zefiro cluster at the INFN computer center in Pisa. Enrico Mazzoni is warmly thanked for assistance.

\end{document}